\documentclass[a4paper,12pt]{article}
\usepackage[utf8]{inputenc}
\usepackage{amssymb}
\usepackage{amsmath}
\usepackage{mathtools}
\usepackage{slashed}
\usepackage{color}
\usepackage{indentfirst}
\usepackage[margin=2.5cm]{geometry}
\usepackage{graphicx}
\usepackage{bbm}
\usepackage{csquotes} 

\renewcommand{\thefootnote}{\fnsymbol{footnote}}

\begin{document}

\title{
\begin{flushright}
\begin{minipage}{0.2\linewidth}
\normalsize
WU-HEP-19-04 \\*[50pt]
\end{minipage}
\end{flushright}
{\Large \bf 
Heterotic Stringy Corrections to Metrics of Toroidal Orbifolds and Their Resolutions
\\*[20pt]}}

\author{Pompey Leung\footnote{
E-mail address: pompeyleung@fuji.waseda.jp
}
\ and\
Hajime~Otsuka\footnote{
E-mail address: h.otsuka@aoni.waseda.jp
}\\*[20pt]
{\it \normalsize 
Department of Physics, Waseda University, 
Tokyo 169-8555, Japan} \\*[50pt]}
\maketitle

\date{
\centerline{\small \bf Abstract}
\begin{minipage}{0.9\linewidth}
\medskip 
\medskip 
\small
We explicitly analyse $O(\alpha')$ corrections 
to heterotic supergravity on toroidal orbifolds and 
their resolutions, which play important roles in 
string phenomenology as well as moduli stabilisation.
Using a conformal factor ansatz that is valid 
only for four dimensional geometries, 
we obtain a closed expression for the $O(\alpha')$ metric corrections in the case of several orbifold limits of K3, namely $T^4/\mathbb{Z}_n$ where $n=2,3,4,6$. However, we find that non-standard embedding requires the inclusion of five-branes on such orbifolds. 
We also numerically investigate the behaviour 
around orbifold fixed points by considering 
the metric correction on the resolution of a 
$\mathbb{C}^2/\mathbb{Z}_2$ singularity. In this case, a non-trivial conformal factor can be obtained in non-standard embedding even without five-branes. 
In the same manner, we generalise our analysis to study metric corrections
on $T^6/\mathbb{Z}_3$ and its resolution described 
by a complex line bundle over $\mathbb{CP}^2$.
Further prospects of utilising these $O(\alpha')$ 
corrected metrics as a novel approach in obtaining 
realistic or semi-realistic Yukawa couplings are discussed.
\end{minipage}
}

\begin{titlepage}
\maketitle
\thispagestyle{empty}
\end{titlepage}

\renewcommand{\thefootnote}{\arabic{footnote}}
\setcounter{footnote}{0}

\tableofcontents

\section{Introduction}

String theory as a promising candidate of quantum gravity predicts higher-derivative corrections known as $\alpha'$-corrections in the low-energy effective action. 
It appears as an expansion parameter in supergravity as a low-energy effective field theory of the superstring, and characterizes the degree to which stringy effects appear in low energy scales relative to the string scale. In essence, $\alpha'$ terms in supergravity provide quantum stringy corrections to the base theory.
  
The presence of $\alpha'$ corrections are significant in string cosmology and phenomenology. In particular, $\alpha'$ corrections play an important role in moduli stabilisation 
represented by the LARGE Volume Scenario (LVS)~\cite{Balasubramanian:2005zx}, where such corrections appear in the K\"ahler potential. 
This is known to occur at $O(\alpha'^3)$ for Type IIB compactifications \cite{Becker:2002nn}, and at $O(\alpha'^2)$ for heterotic compactifications \cite{Ang10}. Thus, heterotic compactifications receive non-trivial $\alpha'$ corrections “earlier” than its Type IIB counterpart. The mechanism by which such non-trivial contributions to the K\"ahler potential appears is due to the breaking of no-scale structure of the potential~\cite{Ang10}. In the heterotic case, although no-scale structure is preserved at $O(\alpha')$, and hence no corrections are generated, it is done so at the cost of shifting the K\"ahler moduli themselves by an $O(\alpha')$ term \cite{Ang10}. Such considerations motivate the study of heterotic $\alpha'$ corrections. Indeed, the results obtained in this paper correspond to explicit expressions for this $O(\alpha')$ shift to the K\"ahler moduli.

On the phenomenological side, models based on toroidal orbifolds and their blow-ups in the context of heterotic compactifications have been well studied \cite{Ibanez:1987sn,Bailin:1987xm,Ibanez:1987pj,Casas:1988hb,Font:1989aj,Hamidi:1986vh,Aspinwall:1994ev,Nib07,Nib08,Dix85,Dixon:1986jc,Kat90}, and physical quantities such as 4D effective Yukawa couplings have subsequently been calculated from them.\footnote{For more details, see e.g. \cite{Bailin:1999nk,Ibanez:2012zz,Blumenhagen:2013fgp} and references therein.} This attraction to toroidal orbifolds and their blow-ups can be traced to the fact that their conformal field theories (CFT) are exact.
However, $\alpha'$ and $g_s$ corrections have not been fully considered so far, and thus calls their derived physical quantities into question. In this regard, by calculating explicit forms of $O(\alpha')$ corrections for toroidal orbifolds and their resolutions, the impact these corrections have on the physical quantities of these models can be further investigated. 

Additionally, in considering heterotic $\alpha'$ corrections, it turns out that $O(\alpha')$ corrections to most 10D heterotic supergravity fields can be expressed in terms of the $O(\alpha')$ correction for the metric, $h_{i\bar{j}}$ \cite{Str86,Ang10}. It thus follows that calculating the metric correction $h_{i\bar{j}}$ is enough to analyse the effects of $O(\alpha')$ corrections. The metric corrections calculated in this paper provide an extension to the usual toroidal orbifold geometries used in model building, and could give rise to new ways to generate 4D effective physical quantities such as the Yukawa coupling.

This paper will be structured as follows: In Section 2, we will review the equations governing the $O(\alpha')$ metric corrections and field corrections. A conformal factor ansatz first found in \cite{Str86} that is only valid in four real dimensions will then be introduced. In Section 3, we will apply this ansatz to several orbifold limits of the K3 surface. By considering fluxes and curvatures localised at orbifold fixed points, an analytic expression for the conformal factor can be found. However, it is argued that the conformal factor at $O(\alpha')$ must necessarily be a constant, otherwise a background torsion must be considered. In Section 4, we will take this one step further and relax the orbifold limit in order to study metric corrections on resolutions of the $\mathbb{C}^2/\mathbb{Z}_2$ singularity. In this case, a numerical solution is obtained and is shown that a non-constant conformal factor can be obtained without the need for five-branes. In Section 5, the same analysis for the conformal factor will be performed in the full six dimensions for the trace of the metric correction $h$, as the previous ansatz is no longer valid. We will find that for the resolution of the $\mathbb{C}^3/\mathbb{Z}_3$ singularity, $h$ obeys the 6D version of the equation of motion for the conformal factor, and behaves similarly with its four dimensional counterpart with the exception that $h$ itself takes negative values due to different boundary conditions. This is however not a problem as the physical meaning of $h$ is the trace of the metric correction, and the trace of the full metric after taking on such a correction remains positive. In Section 6, we summarise and discuss our findings, giving an outlook on several directions to take our results in.

\section{$O(\alpha')$ Corrections to 10D Heterotic Supergravity Fields}

\subsection{The $O(\alpha')$ Metric Correction}

We begin this section by briefly reviewing the $O(\alpha')$ corrections to 10D heterotic supergravity fields, following the exposition given in \cite{Ang10}. We will be working in the same theoretical setting and adopt the following convention for the heterotic effective action in the string frame
\begin{multline} \label{eq:10d action}
    S = \frac{1}{2 \kappa_{10}^2} \int d^{10}x \sqrt{-G} e^{-2 \Phi} 
    \left\{ 
    \mathcal{R}(\omega) + 4(\partial_M \Phi)^2 - \frac{1}{12} H_{MNP}H^{MNP} \right. \\
    \left. - \frac{\alpha'}{8} \big[ \mathrm{tr} \mathcal{F}_{MN}\mathcal{F}^{MN} - \mathcal{R}_{MNAB}(\omega_+) \mathcal{R}^{MNAB}(\omega_+)\big] + O(\alpha'^3) \right\},
\end{multline}
 with indices $M, N, ...$ running from $0$ to $9$, and $\mathrm{tr}$ denoting the trace in the fundamental representation of SO(32). This is related to the trace $\mathrm{Tr}$ in the adjoint representation of the SO(32) and $E_8 \times E_8$ gauge groups by (see e.g. \cite{Candelas:1985en,Green:1987mn})
\begin{equation}
    \mathrm{tr}\mathcal{F}_{MN}\mathcal{F}^{MN} = \frac{1}{30}\mathrm{Tr}\mathcal{F}_{MN}\mathcal{F}^{MN}.
\end{equation}

The 10D fields appearing in Eq.~\eqref{eq:10d action} are the metric $G$, the dilaton $\Phi$, the Neveu-Schwarz (NS) flux $H$, and the Yang-Mills (YM) field strength $\mathcal{F}$. The Ricci scalar $\mathcal{R}(\omega)$ is evaluated with respect to the Levi-Civita connection formed by the metric $G$, whilst the Riemann curvature tensor $\mathcal{R}_{MNAB}(\omega_+)$ is evaluated with respect to the torsion-shifted spin connection
\begin{equation} \label{eq:torshiftcon}
    \omega_+ = \omega + \frac{1}{2} H,
\end{equation}
where the torsion is induced by the $H$-flux. This $H$-flux satisfies the following Bianchi identity coming from the Green-Schwarz anomaly cancellation mechanism \cite{Ang10,Green:1984sg}
\begin{equation} \label{eq:bianchi}
    dH = \frac{\alpha'}{4} \Big[\mathrm{tr} (\mathcal{R_+} \wedge \mathcal{R_+}) - \mathrm{tr} (\mathcal{F} \wedge \mathcal{F})\Big].
\end{equation}
Here, $\mathcal{F}$ is the flux 2-form, and $\mathcal{R}_+$ is the curvature 2-form evaluated with respect to the torision-shifted spin connection Eq.~\eqref{eq:torshiftcon}. 

Rather than working with a warped product geometry with respect to the Einstein frame, going to the string frame allows the 10D spacetime to be expressed as a direct product of a 4D spacetime with a 6D compact manifold $\mathbb{R}^{1,3} \times \mathcal{M}$ even in the presence of non-trivial $H$-flux \cite{Bec06}. Here we make the assumption that the $H$-flux is trivial to leading order and only enters as an $O(\alpha')$ correction. This ensures that the background geometry is K\"ahler, and hence Calabi-Yau to leading order \cite{Str86}.
 Let us also take $\mathcal{M}$ to be Calabi-Yau with a holomorphic vector bundle $E \rightarrow \mathcal{M}$ to accomodate gauge fields.
The $O(\alpha')$ corrections to the 10D heterotic supergravity fields in a complex basis take the following forms \cite{Ang10,Gil03}

\begin{equation} \label{eq:metriccor}
    G_{i\bar{j}} = g_{i\bar{j}} + \alpha' h_{i\bar{j}},
\end{equation}
\begin{equation} \label{eq:dilcor}
    \Phi = \phi_0 - \alpha' \xi h^i_i,
\end{equation}
\begin{equation} \label{eq:3formcor}
    H_{ij\bar{k}} = \alpha' (-\nabla_i h_{j\bar{k}} + \nabla_j h_{i\bar{k}}),
\end{equation}
\begin{equation}
    A_i = A^{(0)}_i + \alpha' A^{(1)}_i,
\end{equation}
\begin{equation}
    \mathcal{F}_{i\bar{j}} = F_{i\bar{j}} + \alpha' F^{(1)}_{i\bar{j}}.
\end{equation}
The $\xi$ appearing in the dilaton correction Eq.~\eqref{eq:dilcor} is a gauge parameter that will be fixed momentarily. The background metric $g_{i\bar{j}}$ is used to raise and lower indices, and the covariant derivatives $\nabla_i$ appearing in Eq.~\eqref{eq:3formcor} are defined with respect to $g_{i\bar{j}}$ as well.

From the above, it is obvious that all fields except the gauge field and its field strength can be expressed in terms of the metric correction $h_{i\bar{j}}$. Calculating $h_{i\bar{j}}$ for certain flux and curvature backgrounds will thus be the prime target of this paper. The equations of motion for $h_{i\bar{j}}$ can be obtained by substituting Eqs.~\eqref{eq:metriccor}, \eqref{eq:dilcor}, and \eqref{eq:3formcor} into the equations of motion for the total metric $G_{MN}$ obtained from Eq.~\eqref{eq:10d action} and collecting terms at $O(\alpha')$
\begin{equation}
    \mathcal{R}_{MN} + 2\nabla_M \nabla_N \Phi - \frac{1}{4} H_{MAB}H_N^{\ AB} - \frac{\alpha'}{4} \Big[ \mathrm{tr} \mathcal{F}_{MP} \mathcal{F}_N^{\ P} - \mathcal{R}_{MPAB} \mathcal{R}_N^{\ PAB} \Big] = O(\alpha'^3).
\end{equation}
This gives us a sourced Lichnerowicz equation for the metric correction, which in a real basis is
\begin{equation}
    \Delta_L h_{mn} + \xi \nabla_m \nabla_n h = \frac{1}{4} \Big[\mathrm{tr} (F_{mp} F_n^{\ p}) - R_{mpqr} R_n^{\ pqr}\Big],
\end{equation}
where the Lichnerowicz operator $\Delta_L$ is given by
\begin{equation}
    \Delta_L h_{mn} = -\frac{1}{2} \nabla^2 h_{mn} - R_{mpnq}h^{pq} + \nabla_{(m} \nabla^p h_{n)p} + R_{p(m}h_{n)}^{p} - \frac{1}{2} \nabla_m \nabla_n h,
\end{equation}
and $h$ is the trace of the metric correction defined as 
\begin{equation}
    h\equiv h^m_m = 2h^i_i.
\end{equation}
Note that $\mathcal{R}$ is constructed using the full metric $G$, while $R$ is constructed using the background metric $g$.

Using the fact that we are working on a Ricci-flat manifold, we can set $R_{mn}=0$. We also gauge away $\xi$ by imposing the following gauge fixing condition\footnote{Note that for $\xi=0$, this is simply the de Donder gauge condition that appears in the analysis of gravitational waves (see e.g. \cite{Car04}).}
\begin{equation}
    \nabla^n h_{mn} = \left( \frac{1}{2} - \xi \right) \nabla_m h.
\end{equation}
 The resulting equation for $h_{mn}$ is the Lichnerowicz equation on a Ricci-flat manifold sourced by background curvatures and fluxes \cite{Bec06},
\begin{equation} \label{eq:genlich}
    -(\nabla^2 \delta^p_m \delta^q_n + 2 R^{p \ q}_{\ m \ n})h_{pq} = \frac{1}{2} \Big[\mathrm{tr} (F_{mp} F^{\ p}_n) - R_{mpqr} R^{\ pqr}_n\Big].
\end{equation}
Furthermore, as a consequence of the Killing spinor equation for the gluino, the flux $F$ appearing in the source is also required to satisfy the Hermitian Yang-Mills equations in order for $\mathcal{N}=1$ supersymmetry to be preserved \cite{Ang10}
\begin{equation} \label{eq:HYM}
    F_{ij} = F_{\bar{i}\bar{j}} = g^{i\bar{j}}F_{i\bar{j}} = 0.
\end{equation}
In general, Eq.~\eqref{eq:genlich} has at most 21 components to solve for in the case of a 3-dimensional complex manifold. Though it is in principle possible to solve such a system of equations, it remains difficult since there are no known analytical expressions for metrics of Calabi-Yau 3-folds.

\subsection{The Conformal Factor Ansatz}

In searching for exact solutions, we turn to the conformal factor ansatz first studied in \cite{Str86} in the construction of supersymmetric vacua with non-vanishing torsion. Under this ansatz, the full metric takes the following form
\begin{equation} \label{eq:confansatz}
	G_{i\bar{j}} = \Omega^2 g_{i\bar{j}}.
\end{equation}
In this way, a conformally Calabi-Yau geometry can be induced by stringy quantum corrections at $O(\alpha')$. Such an ansatz was shown to give non-constant conformal factors only in the case of two-dimensional complex manifolds \cite{Str86}. We shall see shortly that the ansatz Eq.~\eqref{eq:confansatz} allows us to, in a sense, bypass the sourced Lichnerowicz equation Eq.~\eqref{eq:genlich}, and work with a simple Poisson equation. 

To obtain the equation of motion for the conformal factor $\Omega^2$, we use the fact that the NS $H$-flux satisfies the Bianchi identity Eq.~\eqref{eq:bianchi}. Truncating to $O(\alpha')$, we find that we can replace $\mathcal{R_+}$ and $\mathcal{F}$ with $R$ and $F$ respectively. This is possible because $H$ only enters the torsion-shifted spin connection Eq.~\eqref{eq:torshiftcon} at $O(\alpha')$, and so will result in $O(\alpha'^2)$ terms in Eq.~\eqref{eq:bianchi}, namely
\begin{equation} \label{eq:bianchi2}
    dH = \frac{\alpha'}{4} \Big[\mathrm{tr} (R \wedge R) - \mathrm{tr} (F \wedge F)\Big] + O(\alpha'^2).
\end{equation}
In this case, a solution for $H$ can be written in terms of the K\"ahler form $J$ of the internal manifold \cite{Str86}
\begin{equation} \label{eq:Hsol}
    H = i (\partial - \bar{\partial})J.
\end{equation}
Generically, such a non-trivial torsion $H$ is induced at $O(\alpha')$.
Using the fact that on a complex manifold, $2\partial \bar{\partial} = -d (\partial - \bar{\partial})$, we can turn Eq.~\eqref{eq:bianchi2} into a differential equation for $J$, and hence for $g$
\begin{equation} \label{eq:bianchi3}
    -i \partial \bar{\partial}J = \frac{\alpha'}{8} \Big[\mathrm{tr} (R \wedge R) - \mathrm{tr} (F \wedge F)\Big].
\end{equation}
Substituting Eq.~\eqref{eq:confansatz} into Eq.~\eqref{eq:bianchi3} and taking the Hodge dual with respect to the background metric $g$, we obtain a Poisson equation for $\Omega^2$ sourced by flux and curvature terms
\begin{equation} \label{eq:genpoisson}
    \nabla^2 \Omega^2 = \frac{\alpha'}{8} \Big[\mathrm{tr}(R_{mn} R^{mn}) - \mathrm{tr}(F_{mn}F^{mn})\Big].
\end{equation}
Thus, in exchange for considering a restricted class of solutions, one obtains a drastic simplification of Eq.~\eqref{eq:genlich}. Note that we can expand the conformal factor like $\Omega^2 = 1 + \alpha' f + O(\alpha'^2)$. As a result, the $O(\alpha')$ term $fg$ in the full metric Eq.~\eqref{eq:confansatz} turns out to be exactly the shift in the K\"ahler form that is responsible for the breaking of no-scale structure \cite{Ang10}.

It is important to note that the conformal factor ansatz Eq.~\eqref{eq:confansatz} is valid only in four real dimensions; it was shown that only constant $\Omega^2$ can be obtained for $d \neq 4$ dimensions \cite{Str86}. This naturally leads to the K3 surface being chosen as the background geometry, as it is the only non-trivial example of a simply connected, compact Calabi-Yau 2-fold. We thus have to further separate our internal manifold into $\mathcal{M} = \mathcal{M}_4 \times \mathcal{M}_2$, and work with $\mathcal{M}_4$ as the K3 surface. The remaining $\mathcal{M}_2$ is fixed to be a 2-torus $T^2$ for concreteness.

Several conditions must be satisfied in order for Eq.~\eqref{eq:genpoisson} to have solutions \cite{Str86}. In addition to requiring that the gauge field strength $F$ satisfies the Hermitian Yang-Mills equations Eq.~\eqref{eq:HYM}, we also need to respect the following integrability condition\footnote{In general, the right hand side is non-zero due to contributions from the presence of five-branes. Here the right hand side is zero as we are considering vanishing background torsion and we do not consider the existence of five-branes.}
\begin{equation} \label{eq:intcondc2}
    \int_{\mathcal{C}^4} \left[ c_2(F) - c_2(R)\right] = 0,
\end{equation}
where $\mathcal{C}^4$ denotes a 4-cycle. In this way, Eqs.~\eqref{eq:HYM} and \eqref{eq:intcondc2} place constraints on what fluxes and curvatures can source such a conformal factor deformation. Note that the source terms trivially vanish in the standard embedding scenario $F=R$ as mentioned in \cite{Ang10,Str86}. This means that there are no $O(\alpha')$ corrections to the metric.

We would like to mention here that coincidentally, the trace of the metric correction equation Eq.~\eqref{eq:genlich} can be taken to give \cite{Ang10}
\begin{equation} \label{eq:genpoisson2}
    \nabla^2 h = \frac{1}{2} \Big[ R_{mnpq} R^{mnpq} - \mathrm{tr}(F_{mn} F^{mn}) \Big] = R_{i\bar{j}pq} R^{i\bar{j}pq} - \mathrm{tr}(F_{i\bar{j}} F^{i\bar{j}}).
\end{equation}
Though the equations of motion take the same form, the exact relation between the trace of the metric correction $h$ and the conformal factor $\Omega^2$ is not clear at this stage. The only obvious difference is that Eq.~\eqref{eq:genpoisson2} holds for any dimension whilst Eq.~\eqref{eq:genpoisson} only has nontrivial solutions only in four dimensions.

\section{Metric Corrections to $T^4/\mathbb{Z}_n$ Orbifolds}

Having now established that the background geometry of $\mathcal{M}_4$ is the K3 surface, we will proceed to outline its construction. We will be looking at the conformal factor ansatz for the K3 in the orbifold limit in this section, and will develop the analysis further on blow-ups of such orbifolds in Section 4. We aim to examine in more detail the general comments made in \cite{Str86} regarding the feasibility of a non-trivial conformal factor in such a configuration.

\subsection{Orbifold Limits of K3}

The K3 surface can be realised as several different orbifolds $T^4 / \mathbb{Z}_n$, where $n=2,3,4,6$ \cite{Wal88}. Each of these orbifolds is obtained by further identifying points on $T^4$ with rotations corresponding to the discrete symmetry group $\mathbb{Z}_n$. Let us begin by considering the 4-torus as $T^4 = T^2_1 \times T^2_2$ with complex coordinates\footnote{Although we start with two factorisable 2-tori, the discrete group action renders them inseparable, hence the final object is a single, 4-dimensional geometry expressed in the complex coordinates of two factorisable 2-tori.} $z = (z_1, z_2)$. For $n=2, 4$, the tori are given by the complex plane $\mathbb{C}$ identified by $z_i \simeq z_i + 1 \simeq z_i + i, i=1,2$, resulting in a square fundamental region. On the other hand, for $n=3, 6$ the tori are formed by the identification $z_i \simeq z_i + 1 \simeq z_i + e^{\pi i / 3}, i=1,2 $, giving a parallelogram for the fundamental region. In all cases above, the corresponding discrete symmetry group then acts on $T^2_1$ as a rotation by $e^{2 \pi i / n}$, and on $T^2_2$ as a rotation by $e^{-2 \pi i / n}$, where $n$ is the order of $\mathbb{Z}_n$. This generates 16 fixed points for the $T^4 / \mathbb{Z}_2$ orbifold, 9 for the $T^4 / \mathbb{Z}_3$ orbifold, 16 for the $T^4 / \mathbb{Z}_4$ orbifold, and 24 for the $T^4 / \mathbb{Z}_6$ orbifold \cite{Wal88}. 

At this stage, the orbifolds are singular due to the fixed points, and surgery must be performed to remove such singularities. This is done by removing the neighbourhood around each fixed point, and replacing it with an appropriate geometry. In our case, we need a smooth, non-compact Calabi-Yau manifold with $S^3 / \mathbb{Z}_n$ as its boundary at $\infty$. For the $n=2$ case, this manifold is known as an Eguchi-Hanson space, and analogues for other $n$ have been shown to exist \cite{Bec06,Wal88}. For the purposes of this paper, we will denote them collectively as $\mathrm{EH}_n$, Eguchi-Hanson spaces of order $n$. Finally, the orbifold limit of K3 is obtained when the radius of all $\mathrm{EH}_n$ is shrunken to zero.

\subsection{Equation of Motion and Solution}

Having constructed the orbifold limits of K3, we will now proceed to deriving and solving the equation of motion for the conformal factor on them. In the following, we will focus on the case of U(1) flux insertions for concreteness. Solutions to Eq.~\eqref{eq:genpoisson} consistent with the orbifold limit must respect the Hermitian Yang-Mills equations Eq.~\eqref{eq:HYM} and the integrability condition Eq.~\eqref{eq:intcondc2}. In accordance with \cite{Str86}, we thus must take the gauge field strength and the K\"ahler form to only be linear combinations of harmonic (1,1)-forms on the Eguchi-Hanson geometries. In the orbifold limit, the fluxes and curvature thus become localised at the fixed points located at $z_k$ and we have
\begin{equation} \label{eq:deltapoisson}
    \nabla^2 \Omega^2 = \alpha' \pi^2 \sum_{k=1}^{k_n} q_k \delta^{(4)} (z-z_k).
\end{equation}
Here, $k_n$ denotes the total number of fixed points for the $T^4 / \mathbb{Z}_n$ orbifold, and the $q_k$ are “charges” corresponding to the localised fluxes and curvatures on each fixed point. This charge denotes the total contribution of flux and curvature located at each fixed point. The Hermitian Yang-Mills equations Eq.~\eqref{eq:HYM} are satisfied by construction, and the remaining integrability condition Eq.~\eqref{eq:intcondc2} acts as a constraint on the charges. To see this, note that each $q_k$ can be thought of as an integral of $c_2(R) - c_2(F)$ over a small neighbourhood of the fixed point at $z_k$ \cite{Str86}. Integrating over the whole $T^4 / \mathbb{Z}_n$, Eq.~\eqref{eq:intcondc2} becomes the following constraint on the $q_k$
\begin{equation} \label{eq:constraint}
	\sum_{k=1}^{k_n} q_k = 0.
\end{equation}
In the following we shall consider two possible configurations satisfying the above constraint: $q_k\neq0$ and $q_k=0$. We find that, as discussed below, the $q_k\neq0$ case is much harder to treat, requiring higher order $\alpha'$-corrections as Eq.~\eqref{eq:deltapoisson} is insufficient.

\subsection*{Non-Trivial Charges: $q_k\neq0$}

The solution to Eq.~\eqref{eq:deltapoisson} takes advantage of the fact that the source is a sum of delta functions located at each orbifold fixed point. Setting aside the constraint Eq.~\eqref{eq:constraint} for now, let us consider the case of a single delta function source in order to better demonstrate our method. We begin with the following Poisson equation defined on $T^4 / \mathbb{Z}_n$
\begin{equation}
    \nabla^2 \Omega^2 = \alpha' \pi^2 q_k \delta^{(4)} (z-z_k).
\end{equation}
This is solved by the Green's function for the Laplacian defined on $T^4 / \mathbb{Z}_n$. The final solution will thus be a sum of $k_n$ Green's functions each centred at $z_k$ and weighted by their respective charge $q_k$. Now, since the geometry is flat apart from the orbifold singularities, instead of working on the fundamental region of $T^4 / \mathbb{Z}_n$, we can project onto $\mathbb{C}^2$ by rewriting the single delta function source as a lattice of delta functions\footnote{The following procedure used to find the Green's function on $T^4$ is a four-dimensional generalisation of that used in \cite{Mam14,Pol98} to find the Green's function on $T^2$.} spanning all of $\mathbb{C}^2$
\begin{equation}
    \nabla^2 \Omega^2 = \alpha' \pi^2 q_k \sum_{r,s \in \mathbb{Z}} \delta^{(4)} (z-(z_k +r+s\tau)),
\end{equation}
where $\tau=i$ for $n=2,4$ generates a square lattice, and $\tau=e^{\pi i/3}$ for $n=3,6$ generates a parallelogram lattice, corresponding to the fundamental region of each orbifold. Since the Green's function for the Laplacian on $\mathbb{C}^2$ is known to be\footnote{The following are written in compact notation where $z=(z_1, z_2)$ is a vector of the complex variables.}
\begin{equation} \label{eq:C2greens}
    G_{\mathbb{C}^2}(z,\bar{z};z',\bar{z}') = -\frac{1}{4\pi^2} \frac{1}{|z-z'|^2},
\end{equation}
we can apply the Green's function method for a delta function lattice source to obtain the desired solution
\begin{equation}
    \Omega^2(z,\bar{z}) = \alpha' \pi^2 q_k \int_{\mathbb{C}^2} d^4z' \ G_{\mathbb{C}^2}(z,\bar{z};z',\bar{z}') \sum_{r,s\in \mathbb{Z}} \delta^{(4)} (z'-(z_k + r+s\tau)).
\end{equation}
Evaluating this integral, we obtain
\begin{equation} \label{eq:singlesol}
    \Omega^2 (z,\bar{z}) = -\frac{\alpha'}{4} \sum_{r,s \in \mathbb{Z}} \frac{q_k}{|z-(z_k+r+s\tau)|^2}.
\end{equation}
As previously mentioned, the sum of $k_n$ of Eq.~\eqref{eq:singlesol} each weighted by $q_k$ yields the full solution to Eq.~\eqref{eq:deltapoisson}
\begin{equation} \label{eq:orbfinsol}
    \Omega^2 (z,\bar{z}) = -\frac{\alpha'}{4} \sum_{k=1}^{k_n} \sum_{r,s \in \mathbb{Z}} \frac{q_k}{|z-(z_k+r+s\tau)|^2}.
\end{equation}
Justification for the above method is necessary. It is well-known that Green’s functions on compact spaces in fact require an extra constant source term to be consistent \cite{Mam14, Shi00, Pol98}. An intuitive way to understand this is in the context of electrostatics. The electric field due to a point charge on a torus must start and end at distinct points; because it lacks a boundary, the field lines cannot simply extend to infinity. There are two ways in which this is possible: either have a constant background source to absorb whatever comes out of the point charge, or have other charges on the torus cancel it out. In our case, it sufficed to use the ordinary Green’s function defined on $\mathbb{C}^2$ instead of one with an extra term because the integrability condition Eq.~\eqref{eq:intcondc2} automatically ensures that all charges cancel out in the form of Eq.~\eqref{eq:constraint}. It is also easy to see that the solution Eq.~\eqref{eq:orbfinsol} satisfies periodic and $\mathbb{Z}_n$ twisted boundary conditions by construction due to the lattice structure of the source being taken into account when applying the Green's function method.

That being said, the solution Eq.~\eqref{eq:orbfinsol} reveals a rather unsettling result. Because Eq.~\eqref{eq:constraint} must be satisfied, we necessarily must have a combination of both positive and negative $q_k$'s. However, when a $q_k$ is positive, the conformal factor takes negative values in the neighbourhood of the corresponding fixed point. This effectively means that there will be regions around the fixed point that transition from having a positive metric to a negative metric, giving rise to an unphysical region that pinches itself off from the rest of the manifold. In fact, the torsion due to such a conformal factor can be attributed to the presence of five-branes located at these fixed points, in accordance with \cite{Carlevaro:2013vla}. The non-trivial torsion generated by such a configuration must be taken into account when analysing Eq.~\eqref{eq:bianchi2}, which would result in a nonlinear equation of motion for the conformal factor. As such a system is difficult to work with, for now the $q_k=0$ case appears to be a more viable alternative.

\section*{Trivial Charges: $q_k=0$}

The simplest way to satisfy Eq.~\eqref{eq:constraint} is to set all $q_k = 0$ as with the case of standard embedding. This gives us a trivial, i.e., constant conformal factor as the only harmonic functions on a compact manifold are constant. One might wonder whether setting $q_k=0$ is the only possible solution neglecting torsion. One option is to include bulk fluxes only, or in addition to localised fluxes. These would have to cancel out the localised curvature contributions to the source which is positive, since $c_2(R)=24$ for the whole K3, implying that each fixed point has an individual contribution of $3/2$ to $c_2(R)$. As a result, the source for the equation of motion even in the presence of bulk fluxes would include a positive delta function term. From the above argument, it is straightforward to see that we will still end up with negative metric or unphysical regions even if overall fluxes cancel out the localised curvatures; a finite number of bulk fluxes is not enough to “lift" an infinitely deep well due to the positive delta function. 

Thus we conclude that the conformal factor in the orbifold limit of K3 must either be a constant due to vanishing charges $q_k=0$ in standard embedding, or need to be calculated using a non-standard embedding with five-brane configuration which necessarily generates a non-vanishing torsion and gives a nonlinear equation of motion. At first glance this result seems to limit model building possibilities using the conformal factor ansatz as an $O(\alpha')$ correction, however there is another 4D geometry to be considered which may yield fruitful results. This is the resolution of the fixed points of $T^4/\mathbb{Z}_n$, and is the subject of the following section.

\section{Metric Correction to the Resolution of a $\mathbb{C}^2/\mathbb{Z}_2$ Singularity}

The next step in our analysis is to consider relaxing the orbifold limit to approach a smooth K3 by taking the Eguchi-Hanson geometries to have finite size, in other words blowing up the fixed points. As a consequence, information about the metric away from the fixed points are lost, and analysis can only be carried out on local patches around each fixed point. Fortunately, as the metric on such a blow-up is known, we can look at the metric correction on the resolution of a single fixed point. In accordance with the previous section, we will also specify to the case of a U(1) flux, but will also discuss the differences when more U(1)s are included. The key difference between this and the previous geometry is that the delta function source seen on the right hand side of Eq.~\eqref{eq:deltapoisson} now splits into distinct flux and curvature parts as they can now be separately defined on the resolution. We will find that it is precisely the clear distinction of contributions to the source that allow for a non-trivial conformal factor to arise even though the anomaly cancellation condition Eq.~\eqref{eq:intcondc2} is obeyed. This allows us to forego considering the five-brane system in favour of a simple localised flux and curvature configuration.

\subsection{Resolution of Orbifold Singularities}

The explicit metric of such Eguchi-Hanson spaces were first constructed as self-dual solutions to the Euclidean Einstein's equations and are well-known \cite{Bec06,Egu78,Egu80}. In this paper, we will be using a parametrisation of the resolution discussed in \cite{Nib07} and \cite{Nib08}. The construction chosen here describes resolutions of $\mathbb{C}^n / \mathbb{Z}_n$ singularities as complex line bundles over $\mathbb{CP}^{n-1}$~\cite{Calabi} and are uniquely determined by requiring the holonomy to be SU($n$). Following the notation of \cite{Nib07}, we denote these resolutions by $\mathcal{M}^n$. This particular parametrisation of blow-ups readily allows for the inclusion of U(1) fluxes, and is easily generalisable to higher dimensional orbifolds. In this section, we will be looking at the $n=2$ case as it corresponds to the blow-up of $T^4 / \mathbb{Z}_2$ orbifold singularities, though the case of $T^6 / \mathbb{Z}_3$ is also interesting and will be dealt with in the next Section.\footnote{Although $\mathbb{C}^2 / \mathbb{Z}_2$ is non-compact and we are interested in the compact $T^4 / \mathbb{Z}_2$, the results on the $\mathbb{C}^2 / \mathbb{Z}_2$ resolution can be directly applied to $T^4 / \mathbb{Z}_2$ resolutions as we are only considering neighbourhoods around the fixed points, and both are thus indistinguishable.} Unfortunately, this parametrisation of the resolution is only valid for $\mathbb{C}^n/\mathbb{Z}_n$ orbifolds, and adopting it means that we are restricted to looking at only the blow-up of the $T^4 / \mathbb{Z}_2$ orbifold; this analysis does not carry over to orbifolds of other orders. It should also be noted that, this construction is required to be Ricci-flat and thus ensures that the blow-up provides the correct Eguchi-Hanson geometry, i.e., $\mathcal{M}^2 = \mathrm{EH}_2$.

We begin by giving the explicit forms of quantities of interest on the resolution $\mathcal{M}^2$, which is described by a complex line bundle over $\mathbb{CP}^1$. In the following, all quantities are written in terms of the complex coordinates $z$ and $x$. These coordinates are inherited from different coordinate patches on the original $\mathbb{C}^2 / \mathbb{Z}_2$ orbifold during the blow up proceess and correspond to coordinates parametrising the base $\mathbb{CP}^1$, and the fibre $\mathbb{C}$ respectively \cite{Nib08}. The parameter $r$ is used to denote the size of the blow-up; the orbifold limit corresponds to the case when $r \rightarrow 0$.

The K\"ahler metric of the resolution is explicitly written as
\begin{equation} \label{eq:resmetric}
    g = 
    \frac{1}{2} \frac{1}{\sqrt{r+X}}\begin{pmatrix}
        \dfrac{2(r+X)}{\chi^2} + 2|x|^2|z|^2 & \chi x\bar{z}\\[0.75em]
        \chi \bar{x}z & \dfrac{\chi^2}{2} 
    \end{pmatrix},
\end{equation}
where $X=\chi^2 |x|^2$, $\chi = 1+|z|^2$. The metric components in a complex basis are given by $g = g_{i \bar{j}} dz^i d\bar{z}^{\bar{j}} $ for $i = (z, x)$ and $\bar{j}=(\bar{z},\bar{x})$ in that order. This is the new background metric in our analysis, and is the local coordinate patch describing the resolution of a fixed point. It should be stressed that this metric does not describe the region around the singularity itself, but rather an extra geometry that is surgically placed at the neighbourhood around the fixed point in order to achieve an overall smooth geometry. Without loss of generality, let us consider this resolution to be centred around the origin. From the spin connection $\omega$ given in Appendix~\ref{app}, one can construct the curvature 2-form from $R = d\omega + \omega \wedge \omega= R^a_{ \ b i \bar{j}} dz^i \wedge d\bar{z}^{\bar{j}}$. The full form is written as
\begin{multline} \label{eq:M2curvature}
    R = \frac{r}{r+X}  \left[ \begin{pmatrix}
            \dfrac{2}{\chi^2} - \dfrac{2|x|^2|z|^2}{r+X} & -\dfrac{2 \chi^{-1} \bar{x} z}{\sqrt{r+X}} \\[0.75em]
            -\dfrac{2\chi^{-1} x \bar{z}}{\sqrt{r+X}} &
            -\dfrac{2}{\chi^2} + \dfrac{2|x|^2|z|^2}{r+X}
         \end{pmatrix} dz \wedge d\bar{z}
        +\begin{pmatrix}
            -\dfrac{\chi x \bar{z}}{r+X} & -\dfrac{1}{\sqrt{r+X}} \\[0.75em]
            0 &
            \dfrac{\chi x \bar{z}}{r+X}
         \end{pmatrix} dz \wedge d\bar{x} \right. \\[0.5em]
       + \left. \begin{pmatrix}
            -\dfrac{\chi \bar{x} z}{r+X} & 0 \\[0.75em]
            -\dfrac{1}{\sqrt{r+X}} &
            \dfrac{\chi \bar{x} z}{r+X}
         \end{pmatrix} dx \wedge d\bar{z}
        + \begin{pmatrix}
            -\dfrac{1}{2} \dfrac{\chi^2}{r+X} & 0 \\[0.75em]
            0 &
            \dfrac{1}{2} \dfrac{\chi^2}{r+X}
         \end{pmatrix} dx \wedge d\bar{x} \right].
\end{multline}
Though lengthy, the individual components of $R$ will be needed to compute the curvature contribution to the source on the right hand side of Eq.~\eqref{eq:genpoisson}. It is easy to check that Eq.~\eqref{eq:M2curvature} is traceless in accordance with the SU(2) holonomy of $\mathcal{M}^2$ \cite{Nib07}.

The final ingredient needed for the equation of motion is the background U(1) flux. SO(32) and $E_8 \times E_8$ each admits up to 16 U(1) fluxes given by the following expression
\begin{equation} \label{eq:fluxembed}
    iF_V = iF V^I H_I,
\end{equation}
where $V^I$ and $H_I$ for $I=1,... ,16$ are the gauge shift vectors and generators of the Cartan subgroup respectively. The shift vectors $V^I$ can contain only integer or only half-integer entries. The U(1) field strength $F$, is conveniently constructed by taking an already existing U(1) connection 1-form on the base $\mathbb{CP}^1$ and extending it to satisfy the Hermitian Yang-Mills equations, i.e., to preserve $\mathcal{N}=1$ supersymmetry. $F$ is defined by this extended U(1) connection and takes the following form
\begin{multline} \label{eq:M2flux}
    iF = - \sqrt{\frac{r}{r+X}} \left[ \left( \frac{1}{\chi^2} - \frac{|x|^2|z|^2}{r+X}\right) dz \wedge d\bar{z} - \frac{1}{2}\frac{\chi\bar{x}z}{r+X} dx \wedge d\bar{z} \right. \\
      \left. - \frac{1}{2}\frac{\chi x \bar{z}}{r+X} dz \wedge d\bar{x} - \frac{1}{4} \frac{\chi^2}{r+X} dx \wedge d\bar{x} \right],
\end{multline}
where the components of $iF$ can be read off from $iF=iF_{i \bar{j}}\  dz^i \wedge d\bar{z}^{\bar{j}}$. By a simple calculation, it can be checked that Eq.~\eqref{eq:M2flux} satisfies the Hermitian Yang-Mills equations Eq.~\eqref{eq:HYM}.

As with Section 3 however, we still have to make sure the integrability condition Eq.~\eqref{eq:intcondc2} is satisfied by both the curvature and background fluxes, i.e., that the integrated Bianchi identity vanishes on $\mathcal{M}^2$. Let us start by working out the second Chern classes $c_2(F)$ and $c_2(R)$. For concreteness, we will work with SO(32) in the following. To extend this analysis to $E_8 \times E_8$, see Appendix \ref{app}. Using the normalisation $\mathrm{tr}(H_I H_J) = \delta_{IJ}$, we find that
\begin{equation}
    \mathrm{tr}(iF_V \wedge iF_V) = (iF\wedge iF) V^I V^J \mathrm{tr}(H_I H_J) = (V^I)^2 iF \wedge iF.
\end{equation}
The second Chern classes on $\mathcal{M}^2$ can then be worked out straightforwardly to give
\begin{equation} \label{eq:chern2F}
    c_2(F) = -\frac{1}{2}\frac{1}{(2\pi i)^2} \mathrm{tr} (iF_V \wedge iF_V) = \frac{1}{(2\pi i)^2} \frac{(V^I)^2r}{4(r+X)^2}\ dz \wedge d\bar{z} \wedge dx \wedge d\bar{x},
\end{equation}
\begin{equation} \label{eq:chern2R}
    c_2(R) = -\frac{1}{2}\frac{1}{(2\pi i)^2} \mathrm{tr} (R \wedge R) = \frac{1}{(2\pi i)^2} \frac{3r^2}{(r+X)^3}\ dz \wedge d\bar{z} \wedge dx \wedge d\bar{x}.
\end{equation}
By using the following identity on $\mathcal{M}^2$
\begin{equation} \label{eq:identity}
   \frac{1}{(2\pi i)^2} \int_{\mathcal{M}^2} \frac{dz \wedge d\bar{z} \wedge dx \wedge d\bar{x}}{(r+X)^p} = \frac{1}{p-1} \frac{1}{r^{p-1}},
\end{equation}
which holds for $p>1$, we can show that
\begin{equation} \label{eq:chern2Fint}
    \int_{\mathcal{M}^2} c_2(F) = \frac{(V^I)^2}{4},
\end{equation}
\begin{equation} \label{eq:chern2Rint}
    \int_{\mathcal{M}^2} c_2(R) = \frac{3}{2}. 
\end{equation}
Thus, for the integrated Bianchi identity to vanish, we must have
\begin{equation} \label{eq:M2intcond}
    \sum_{I=1}^{16} (V^I)^2 = 6.
\end{equation}
In this way, the integrability condition Eq.~\eqref{eq:intcondc2} has yet again manifested in the form of a constraint characterising allowed source configurations, this time on the possible values the gauge shift vectors $V^I$ can take. Though in general we can consider up to 16 shift vectors at once in Eq.~\eqref{eq:M2intcond}, for concreteness we will consider a single background U(1) flux, and thus require a single shift vector to satisfy $V^2 = 6$. We will later see that Eq.~\eqref{eq:M2intcond} allows us to ensure that the anomaly cancellation condition is satisfied without having to use five-branes or consider bulk fluxes, at the same time giving us some degree of freedom in choosing the flux configuration.

\subsection{Equation of Motion and Solution}
Armed with the background metric $g$, the curvature 2-form $R$, and the U(1) flux $F$, we are now ready to compute the equation of motion Eq.~\eqref{eq:genpoisson} for the conformal factor on the resolution of the orbifold singularity. The Laplacian on $\mathcal{M}^2$ is highly non-trivial compared to the orbifold case. Using the metric above, it is calculated to be

\begin{equation} \label{eq:laplacian}
    \nabla^2 \Omega^2 = \bigg[ \frac{\chi^2}{\sqrt{r+X}}\partial_z \partial_{\bar{z}} - \frac{2\chi z \bar{x}}{\sqrt{r+X}} \partial_z \partial_{\bar{x}} - \frac{2\chi \bar{z} x}{\sqrt{r+X}} \partial_{\bar{z}} \partial_x
    + \left( \frac{4 \sqrt{r+X}}{\chi^2} + \frac{4 |x|^2 |z|^2}{\sqrt{r+X}}\right) \partial_x \partial_{\bar{x}} \bigg] \Omega^2
    .
\end{equation}
Similarly, the source can be calculated by taking the trace of the curvature and flux 2-forms to give
\begin{equation}
    \rho(|z|,|x|) = \frac{\alpha'}{8}\left[ -\frac{12r}{(r+X)^2}  + \frac{24 r^2}{(r+X)^3} \right],
\end{equation}
where the first and second terms are the flux and curvature contributions respectively. 

It should be pointed out here that the sources can be expressed in terms of orbifold delta functions defined on $\mathcal{M}^2$ for the curvature and flux; using Eqs.~\eqref{eq:chern2Fint} and \eqref{eq:chern2Rint} one is able to construct top forms that exhibit delta function-like behaviour from the curvature and flux 2-forms. These are \cite{Nib07,Nib08}
\begin{equation}
    \delta_{r,R} (|z|,|x|)\ dz \wedge d\bar{z} \wedge dx \wedge d\bar{x} = \frac{1}{12\pi^2} \mathrm{tr} (R \wedge R) = -\frac{1}{2\pi^2} \frac{r^2}{(r+X)^3}\ dz \wedge d\bar{z} \wedge dx \wedge d\bar{x},
\end{equation}
\begin{equation}
    \delta_{r,F} (|z|,|x|)\ dz \wedge d\bar{z} \wedge dx \wedge d\bar{x} = -\frac{1}{12\pi^2} \mathrm{tr} (iF_V \wedge iF_V) = -\frac{1}{4\pi^2} \frac{r}{(r+X)^2}\ dz \wedge d\bar{z} \wedge dx \wedge d\bar{x}.
\end{equation}
These orbifold delta functions become increasingly peaked around the origin in the limit $r \rightarrow 0$, and are equal to one when integrated over the resolution $\mathcal{M}^2$
\begin{equation}
   \int_{\mathcal{M}^2} \delta_{r,R} (|z|,|x|)\ dz \wedge d\bar{z} \wedge dx \wedge d\bar{x} = \int_{\mathcal{M}^2} \delta_{r,F} (|z|,|x|)\ dz \wedge d\bar{z} \wedge dx \wedge d\bar{x} =  1.
\end{equation}
We can now rewrite our sources in terms of these delta functions as
\begin{equation} \label{eq:M2source}
    \rho(|z|,|x|) = -48 \alpha' \pi^2 [\delta_{r,F} (|z|,|x|) - \delta_{r,R} (|z|,|x|)].
\end{equation}
Thus we see how in the blow-up limit, the contributions to the delta function sources in Eq.~\eqref{eq:deltapoisson} can be separated into distinct curvature and flux contributions, each in terms of their respectively defined orbifold delta functions. This is analogous to the delta function sources in the orbifold limit, except for the fact that information about curvature and flux contributions are combined into the $q_k$ in the $r\rightarrow0$ limit. Because the equation of motion comes from taking the Hodge dual of the Bianchi identity Eq.~\eqref{eq:bianchi3}, we expect that the source Eq.~\eqref{eq:M2source} integrated over $\mathcal{M}^2$ also vanishes due to the vanishing integrated Bianchi identity Eq.~\eqref{eq:intcondc2}. It is straightforward to demonstrate this as a consistency check now that the source is written in terms of the orbifold delta functions $\delta_{r,R}$ and  $\delta_{r,F}$; integrating Eq.~\eqref{eq:M2source} over $\mathcal{M}^2$ immediately gives zero. We would like to emphasise here that Eq.~\eqref{eq:M2source} tells us that the source consists of localised fluxes and curvatures just like in the orbifold limit; it is not necessary to consider bulk fluxes in such a setting. 

The full equation of motion currently looks as such
\begin{multline} \label{eq:bigpoisson}
    \bigg[ \frac{\chi^2}{\sqrt{r+X}}\partial_z \partial_{\bar{z}} - \frac{2\chi z \bar{x}}{\sqrt{r+X}} \partial_z \partial_{\bar{x}} - \frac{2\chi \bar{z} x}{\sqrt{r+X}} \partial_{\bar{z}} \partial_x \\
    + \left( \frac{4 \sqrt{r+X}}{\chi^2} + \frac{4|x|^2 |z|^2}{\sqrt{r+X}}\right) \partial_x \partial_{\bar{x}} \bigg] \Omega^2 
    = \frac{\alpha'}{8}\left[-\frac{12r}{(r+X)^2} + \frac{24r^2}{(r+X)^3}\right].
\end{multline}
At first glance it would appear to be rather difficult to solve a Poisson equation with such a complicated Laplacian operator. The problem at hand however can actually be reduced to a 2D Poisson equation by taking advantange of the fact that $\mathcal{M}^2$ possesses a $U(1)_{\mathbb{CP}^1} \times U(1)_{\mathbb{C}}$ symmetry, i.e., it is invariant with respect to separate rotations on the base $\mathbb{CP}^1$ and the fibre $\mathbb{C}$. This is easily seen by rewriting Eq.~\eqref{eq:resmetric} in polar coordinates for $z$ and $x$
\begin{align}
    z &= |z| e^{i \theta_z}, & x &= |x| e^{i \theta_x}.
\end{align}
In this basis, the resulting real metric of $\mathcal{M}^2$ is
\begin{equation} \label{eq:polarresmetric}
    g = 
    \frac{1}{2}\frac{1}{\sqrt{r+X}}\begin{pmatrix}
        \dfrac{2(r+X)}{\chi^2} + 2|x|^2|z|^2 & 0 & \chi |x||z| & 0 \\[0.75em]
        0 & \left[ \dfrac{2(r+X)}{\chi^2} + 2|x|^2|z|^2 \right]|z|^2 & 0 & \chi |x|^2|z|^2 \\[0.75em]
        \chi |x||z| & 0 & \dfrac{\chi^2}{2} & 0 \\[0.75em]
        0 & \chi |x|^2|z|^2 & 0 & \dfrac{\chi^2|x|^2}{2}
    \end{pmatrix},
\end{equation}
 where the metric components in polar coordinates are given by $g=g_{mn} dx^m dx^n$ with indices running in the order $m,n=(|z|, \theta_z, |x|, \theta_x)$. It is then obvious by inspection that the real metric Eq.~\eqref{eq:polarresmetric} is independent of the angular variables $\theta_z$ and $\theta_x$ since $\chi=\chi(|z|)$ and $X=X(|z|,|x|)$. We can immediately conclude that $\mathcal{M}^2$ has a $U(1)_{\mathbb{CP}^1} \times U(1)_{\mathbb{C}}$ isometry generated by the Killing vectors $\partial_{\theta_z}$ and $\partial_{\theta_x}$. Apart from the metric, notice also from the right hand side of Eq.~\eqref{eq:bigpoisson} that the source is rotationally symmetric as it depends only on $X$. The angular independence of both the metric and source suggests that we should seek out a rotationally invariant solution\footnote{See Appendix~\ref{app2} for more details.} to Eq.~\eqref{eq:bigpoisson}. Neglecting $\partial_{\theta_z}$ and $\partial_{\theta_x}$ terms results in the reduced form of the equation of motion for the conformal factor on $\mathcal{M}^2$
\begin{multline} \label{eq:polarpoisson}
	    \bigg[ \frac{\chi^2}{\sqrt{r+X}}\left( \partial_{|z|}^2+\frac{1}{|z|}\partial_{|z|} \right) - \frac{4\chi |z| |x|}{\sqrt{r+X}} \partial_{|z|} \partial_{|x|}\\
	    + \left( \frac{4 \sqrt{r+X}}{\chi^2} + \frac{4|x|^2 |z|^2}{\sqrt{r+X}}\right) \left( \partial_{|x|}^2+\frac{1}{|x|}\partial_{|x|} \right) \bigg] \Omega^2 (|z|,|x|)
    = \frac{\alpha'}{8}\left[-\frac{12r}{(r+X)^2} + \frac{24r^2}{(r+X)^3}\right].
\end{multline}

Though we have managed to reduce the original 4D problem to a 2D one, the mixed derivative between the $|z|$ and $|x|$ coordinates still makes it difficult to find an analytical solution. We will thus numerically solve this differential equation using Mathematica in order to study the behaviour of the conformal factor for different values of the blow-up parameter $r$.

Let us now deduce the boundary conditions for both $|z|$ and $|x|$ appropriate to our problem.
Recall that $z$ is the local complex coordinate describing $\mathbb{CP}^1 \simeq S^2$. Let us take the origin of $z$ to be at one of the poles, say the south pole of $\mathbb{CP}^1$. For our solution to be a continuous function on $\mathbb{CP}^1$ it must join together smoothly at the poles, thus it is easy to see that we should prescribe Neumann boundary conditions at $|z|=0$ and $|z| = \infty$ which correspond to the south and north poles respectively. For $|z|=\infty$, one can also understand this condition by flipping the poles via the coordinate transformation $w=1/z$. Since continuity implies the Neumann condition must hold at $|w|=0$, it must also hold at $|z|=\infty$. From these considerations, we adopt the following Neumann boundary conditions for $|z|$
\begin{equation} \label{eq:zBC}
    \left. \frac{\partial \Omega^2}{\partial |z|} \right|_{|z|=0} = \left. \frac{\partial \Omega^2}{\partial |z|} \right|_{|z|= \infty} = 0.
\end{equation}
As for the coordinate $x$ which parametrises the complex line over $\mathbb{CP}^1$, the Neumann boundary conditions at $|x|=0$ are also due to the continuity requirement, this time arising from the choice of polar coordinates. Since $\mathcal{M}^2$ is asymptotically flat as $|x|\rightarrow \infty$, this corresponds to being far away from the singularity. From the construction of the orbifold limits of K3, this means that the resolution $\mathcal{M}^2$ is surgically connected to the bulk K3 in the region $|x|\rightarrow \infty$. Because the curvature and flux source terms are localised around the fixed point even in the blow up (see the right hand side of Eq.~\eqref{eq:polarpoisson}), we assume that the effects of the source cannot be felt far away from the fixed point. Thus we assume that the deformation due to the conformal factor $\Omega^2$ at $|x|\rightarrow \infty$ is trivial, hence we arrive at the following boundary conditions for $|x|$
\begin{align} \label{eq:xBC}
    \left. \frac{\partial \Omega^2}{\partial |x|} \right|_{|x|=0} &= 0, &
    \Omega^2(|z|,|x|\rightarrow\infty) &= 1.
\end{align}
For computational purposes however, we impose boundary conditions for $|z|$ and $|x|$ along a sufficiently large, but finite radius $L$ instead of at infinity to obtain a numerical approximation. This can be justified by the fact that the source is localised around the origin and tends to 0 as both $|z|,|x| \rightarrow \infty$.

\begin{figure}[htbp]
  \begin{minipage}{0.3\hsize}
   \begin{center}
     \includegraphics[width=50mm]{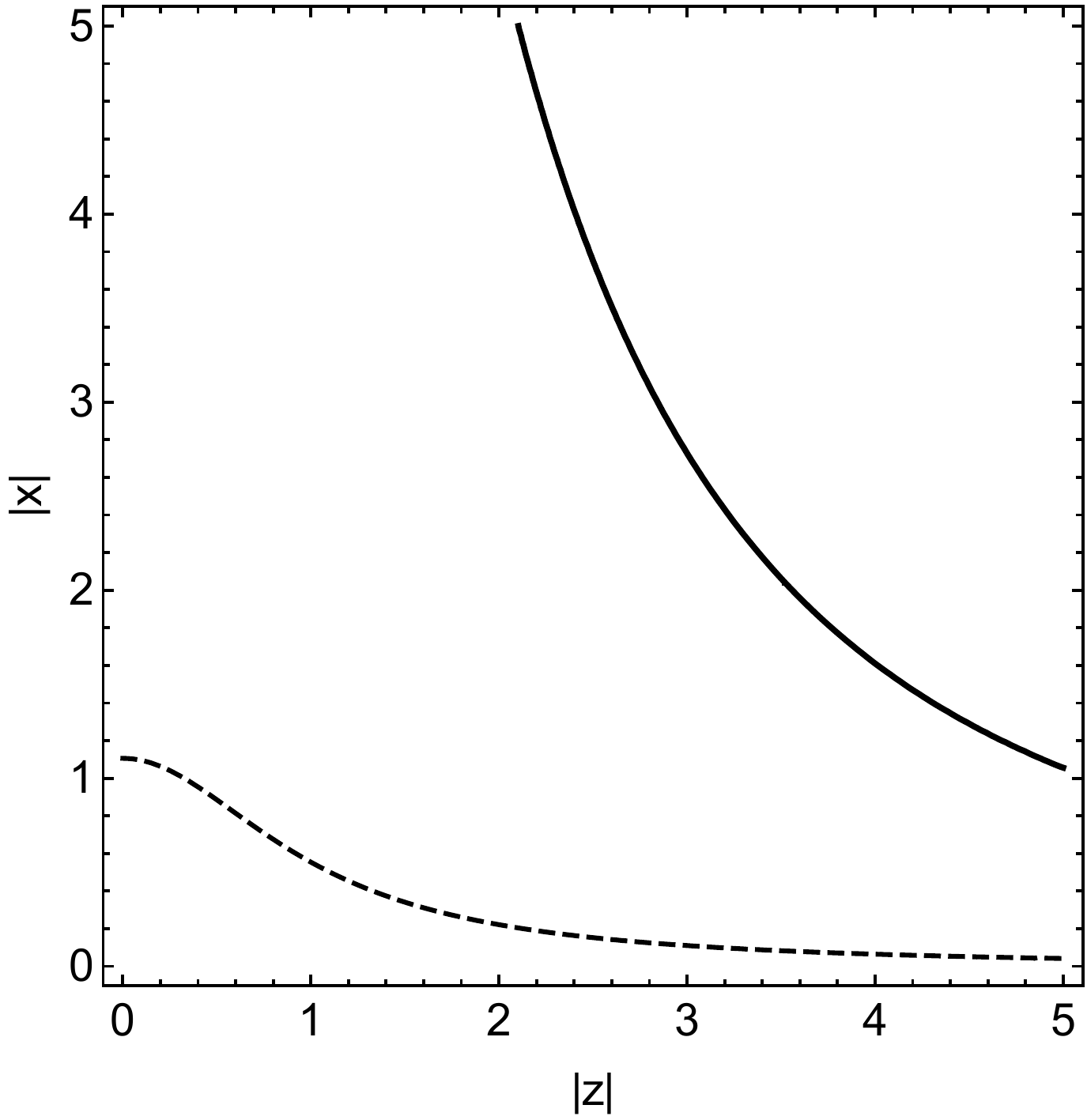}
    \end{center}
   \end{minipage}
  \begin{minipage}{0.3\hsize}
   \begin{center}
    \includegraphics[width=50mm]{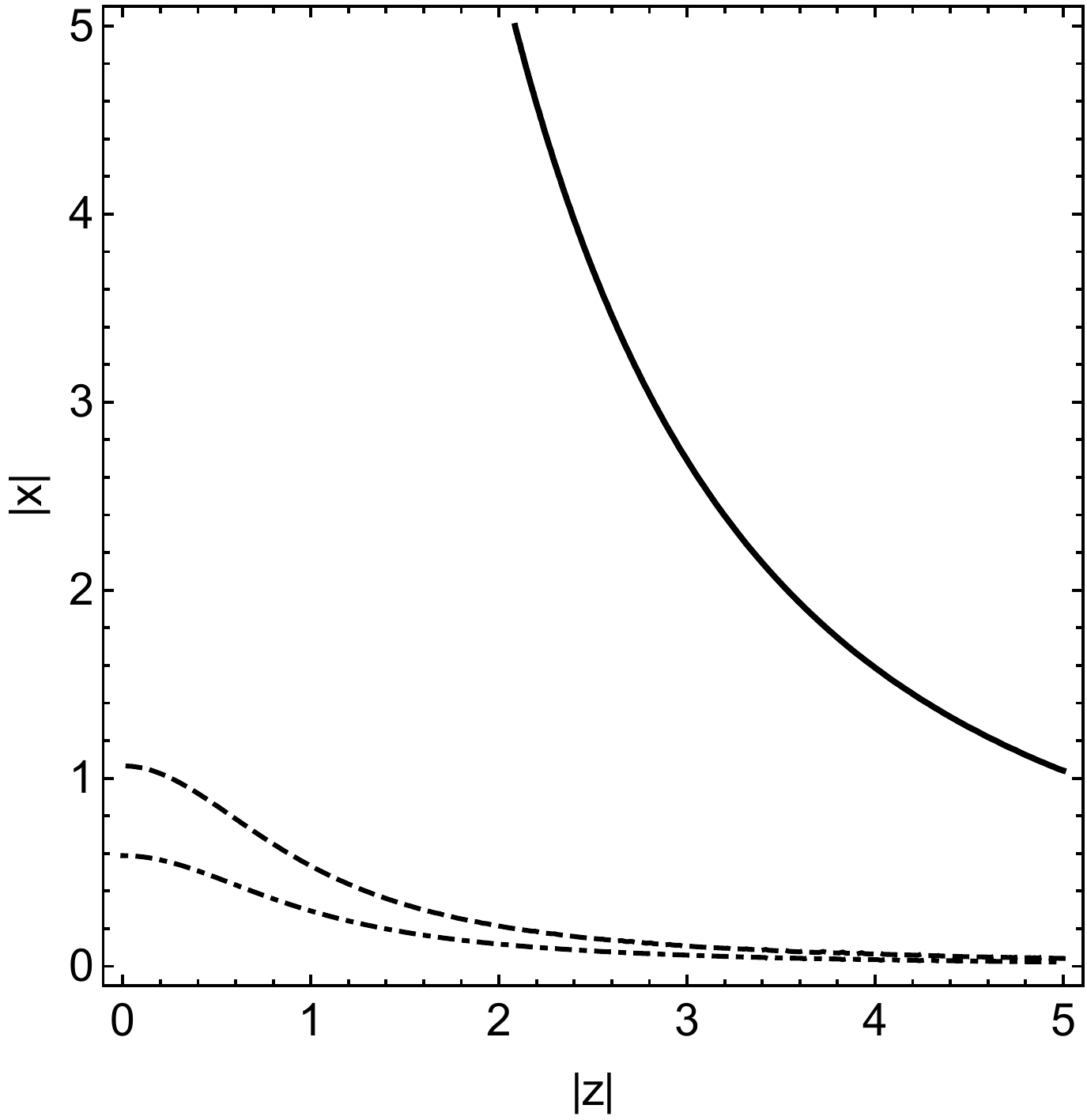}
   \end{center}
  \end{minipage}
  \begin{minipage}{0.3\hsize}
   \begin{center}
    \includegraphics[width=50mm]{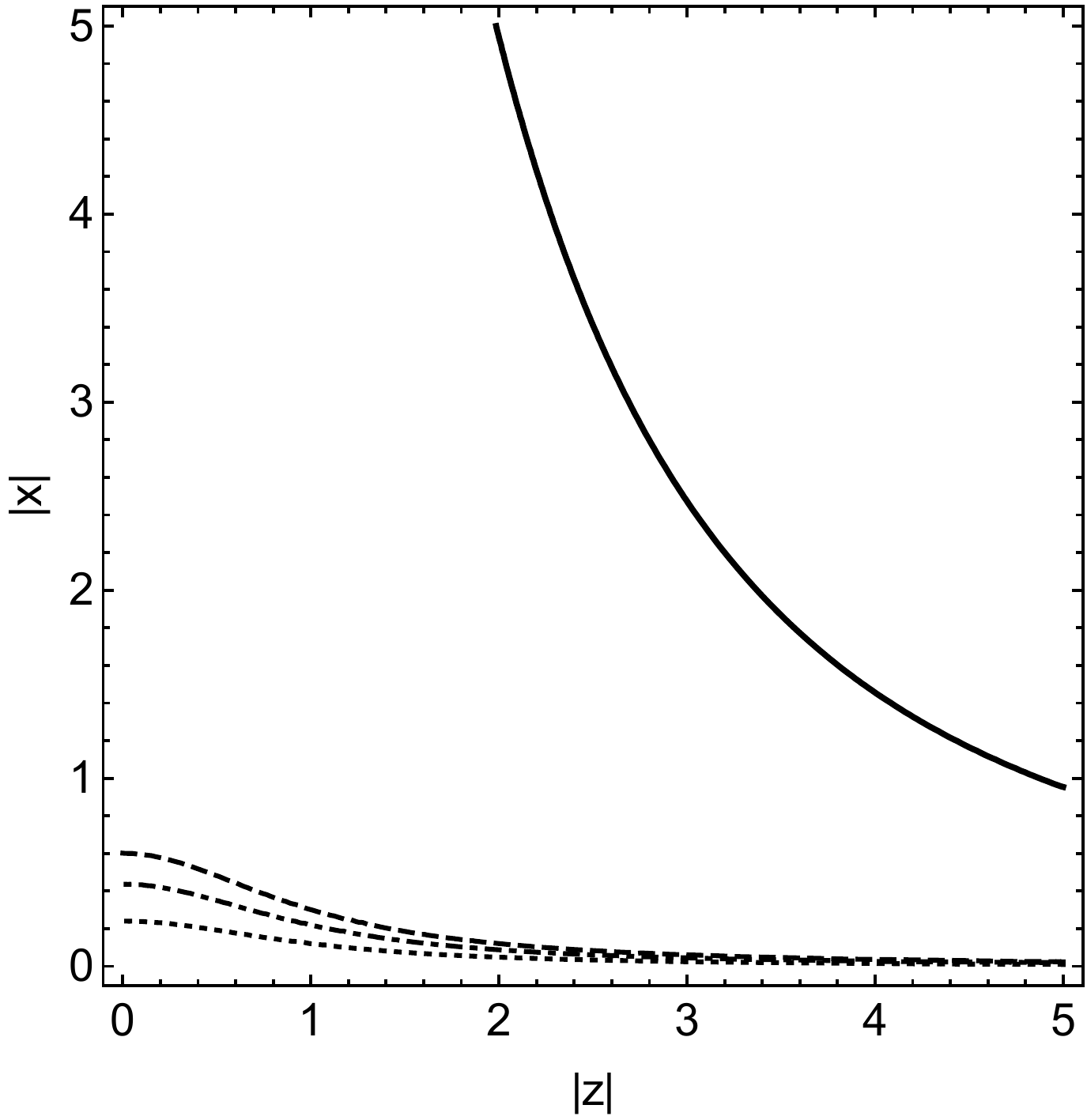}
   \end{center}
  \end{minipage}
    \caption{Numerical solution for the conformal factor $\Omega^2$ on the resolution $\mathcal{M}^2$ on the $(|z|, |x|)$-plane, where the blow-up parameter is chosen as $r=5, 1, 0.1$ from left to right. 
    In all panels, the solid, dashed, dot-dashed and dotted curve denotes $\Omega^2 = 1, 0.99, 0.98, 0.95$, respectively.}
    \label{fig:4D_r=5_1_01}
\end{figure}

Fig.~\ref{fig:4D_r=5_1_01} represents plots for the numerical solution to Eq.~\eqref{eq:polarpoisson} for blow-up parameters $r=5, 1$, and $0.1$, with $\alpha' = 1/2$ as in the orbifold limit case. Due to computational limitations, here we have set $L=10$ as the asymptotically far away “boundary".
We can see from Fig.~\ref{fig:4D_r=5_1_01} that the conformal factor $\Omega^2$ is 1 far away from the origin, but gradually decreases as we approach the origin. The conformal factor at the origin takes a minimum value somewhere between 0.9 and 1 for all $r$ plotted in Fig.~\ref{fig:4D_r=5_1_01}, with this minimum getting smaller as $r$ decreases, resulting in a steeper slope in the neighbourhood of the origin for decreasing $r$. This is reflected in the appearance of more contour lines for smaller $r$, and suggests that the warping of the background metric is negligible far away from the origin, but is scaled down to around 0.9 times the background metric near the centre of the fixed point where the flux and curvature sources are localised. 

Based on this numerical solution, we can conclude that on the resolution of a $T^4/\mathbb{Z}_2$ orbifold fixed point, namely $\mathcal{M}^2$, a non-constant conformal factor can be achieved even in the non-standard embedding scenario without five-branes given that the integrated Bianchi identity Eq.~\eqref{eq:M2intcond} is satisfied. This is in stark contrast to the $T^4/\mathbb{Z}_n$ orbifold limits of K3 which requires the inclusion of five-branes in the analysis when considering non-standard embedding.

\section{Metric Correction to the Resolution of a $\mathbb{C}^3/\mathbb{Z}_3$ Singularity}
Compared to the $T^4/\mathbb{Z}_2$ orbifold, 6D toroidal orbifold models are more phenomenologically attractive to derive the spectra of the Standard Model. 
To understand $O(\alpha')$ corrections to 6D orbifold models, in this section, we focus on the $T^6/\mathbb{Z}_3$ orbifold and its resolution described by the resolution of a $\mathbb{C}^3/\mathbb{Z}_3$ singularity as a complex line bundle over $\mathbb{CP}^2$. Note that as shown in Eq.~\eqref{eq:genpoisson2}, the trace of the generic (not the conformal factor ansatz) metric correction given in \cite{Ang10} has the same equation of motion as the conformal factor, but is not restricted to four real dimensions. 
This means that as a first step to considering more general forms of the correction, we can study the 6D case but only for the trace of the metric.\footnote{It is straightforward to analyse all the elements of the deformed metric employing the following method, because we know the background metric itself. The difficulty lies in the fact that we will have to solve a system of 21 coupled partial differential equations in order to understand these generic deformations.} 
In the following section, we analyse the $\alpha'$ corrections to the metric of the resolution of a $T^6/\mathbb{Z}_3$ orbifold.  Let us briefly comment on the correction in the case of the orbifold limit of $T^6/\mathbb{Z}_3$. This orbifold limit has 27 fixed points and the correction to the trace of the metric $h$ of the $T^6/\mathbb{Z}_3$ orbifold has the same form as Eq.~\eqref{eq:orbfinsol}. However, we need to take care to use the $\mathbb{C}^3$ Green's function in solving over $T^6/\mathbb{Z}_3$ which behaves like $-1/|z|^4$, 
\begin{equation} \label{eq:C3greens}
    G_{\mathbb{C}^3}(z,\bar{z};z',\bar{z}') = -\frac{1}{4\pi^3} \frac{1}{|z-z'|^4},
\end{equation}
and to replace $z_k$ with the 27 fixed points of $T^6/\mathbb{Z}_3$, 
\begin{equation} \label{eq:T6orbsol}
    h (z,\bar{z}) = -\frac{1}{4\pi^3} \sum_{k=1}^{27} \sum_{r,s \in \mathbb{Z}} \frac{q_k}{|z-(z_k+r+s\tau)|^4}.
\end{equation}
In this case then, we also see the same issue that plagued the $T^4/\mathbb{Z}_2$ orbifold in the non-standard embedding scenario - that positive $q_k$ will lead to a negative metric correction around certain fixed points that cannot be lifted because they diverge to $-\infty$ towards the centre of the fixed point.

\subsection{Resolution of Orbifold Singularities}
As discussed in Section 4, the resolution of $T^6/\mathbb{Z}_3$ is described by the resolution of a $\mathbb{C}^3/\mathbb{Z}_3$ singularity as complex line bundle over $\mathbb{CP}^2$, where the K\"ahler metric has the following form \cite{Nib07}
\begin{align} \label{eq:resmetricT6}
    g = 
    \frac{1}{(r+X)^{2/3}}
    \begin{pmatrix}
        \dfrac{(r+X)(1+|z_2|^2)+|x|^2|z_1|^2\chi^3}{\chi^2}  
        & -\dfrac{z_1\bar{z}_2(r+X-|x|^2\chi^3)}{\chi^2}
        & \dfrac{\chi^2 \bar{x}z_1}{3}\\[0.75em]
        -\dfrac{\bar{z}_1z_2(r+X-|x|^2\chi^3)}{\chi^2}
        & \dfrac{(r+X)(1+|z_1|^2)+|x|^2|z_2|^2\chi^3}{\chi^2}
        & \dfrac{\chi^2 \bar{x}z_2}{3}\\[0.75em]
        \dfrac{\chi^2 x\bar{z}_1}{3}
        & \dfrac{\chi^2 x\bar{z}_2}{3}
        & \dfrac{\chi^3}{9}
    \end{pmatrix}
    ,
\end{align} 
with $X=\chi^3|x|^2$ and $\chi=1+|z_1|^2+|z_2|^2$. 
Here, $z_1$, $z_2$, and $x$ denote the coordinates of base $\mathbb{CP}^2$ and the fibre $\mathbb{C}$ respectively and $r$ parametrises the size of the blow-up. Using the above K\"ahler metric $g_{i\bar{j}}$ in a complex basis $i=(z_1,z_2,x)$ and $\bar{j}=(\bar{z}_1, \bar{z}_2, \bar{x})$, one can calculate the curvature 2-form $R=d \omega + \omega \wedge \omega = R^a_{\ b i \bar{j}} dz^i \wedge d\bar{z}^{\bar{j}}$ as shown in Appendix~\ref{app}. In a similar manner to the resolution of a $\mathbb{C}^2/\mathbb{Z}_2$ singularity, this curvature tensor is also traceless due to the SU(3) holonomy of $\mathcal{M}^3$. 

To satisfy the anomaly cancellation condition Eq.~\eqref{eq:intcondc2}, we require the existence of a background field strength. In our 6D case, the 4-cycle ${\cal M}$ in Eq.~\eqref{eq:intcondc2} corresponds to the base $\mathbb{CP}^2$ and the submanifold ${\cal M}^2$ formed by setting $z_1=0$. To simplify our analysis, 
we consider a background U(1) gauge flux, where the U(1) connection 1-form is obtained by taking the trace\footnote{See Appendix \ref{app} for the detailed construction of the background U(1) flux.} of a U(2) 1-form on $\mathbb{CP}^2$.
The U(1) gauge flux satisfying the Hermitian Yang-Mills equations takes the following form
\begin{equation}
\begin{split}
    iF &= - \left(\frac{r}{r+X}\right)^{2/3} \biggl[ \left(\frac{1+|z_2|^2}{\chi^{2}} - \frac{2\chi^3|x|^2|z_1|^2}{r+X}\right) dz_1 \wedge d\bar{z}_1 
      -\frac{z_1\bar{z}_2}{\chi^2}\left(1+\frac{2\chi^3|x|^2}{r+X}\right)
      dz_2\wedge d\bar{z}_1\biggl.\\
    & -\frac{z_1\bar{z}_2}{\chi^2}\left(1+\frac{2\chi^3|x|^2}{r+X}\right)
      dz_1\wedge d\bar{z}_2  
      +\left(\frac{1+|z_1|^2}{\chi^{2}} - \frac{2\chi^3|x|^2|z_2|^2}{r+X}\right) dz_2 \wedge d\bar{z}_2
      -\frac{2\chi^2 x\bar{z}_1}{3(r+X)}dz_1\wedge d\bar{x}\\
    &  -\frac{2\chi^2 x\bar{z}_2}{3(r+X)}dz_2\wedge d\bar{x}
      -\frac{2\chi^2 z_1\bar{x}}{3(r+X)}dx\wedge d\bar{z}_1
      -\frac{2\chi^2 z_2\bar{x}}{3(r+X)}dx\wedge d\bar{z}_2
      -\frac{2\chi^3}{9(r+X)}dx\wedge d\bar{x}
       \biggl],
\end{split}
\end{equation}
which is quantized on 2-cycles of ${\cal M}^3$ and inserted into the Cartan direction $H_I$ of $E_8\times E_8$ or SO(32) as with Eq.~\eqref{eq:fluxembed}. The components are again read from $iF_{i \bar{j}}\  dz^i \wedge d\bar{z}^{\bar{j}}$ as in the $\mathcal{M}^2$ case.
We note that the anomaly cancellation condition Eq.~\eqref{eq:intcondc2} constrains 
the oribifold shift vector in a similar manner to the $\mathcal{M}^2$ case by \cite{Nib07}
\begin{equation} \label{eq:M3intcond}
    \sum_{I=1}^{16} (V^I)^2 = 12.
\end{equation}
In the following analysis, we again specify to one background U(1) flux and impose the constraint $V^2=12$.

\subsection{Equation of Motion and Solution}
Let us compute the $O(\alpha')$ correction to the trace of the metric which obeys the same equation of motion as the conformal factor, namely
\begin{align} \label{eq:EOMT6}
    g^{i\bar{j}}\partial_i\partial_{\bar{j}} h=\rho_F(|z_1|,|z_2|,|x|)+\rho_R(|z_1|,|z_2|,|x|),
\end{align}
where
\begin{align} \label{eq:resmetricT6inv}
    g^{-1} = 
    \frac{1}{(r+X)^{1/3}}
    \begin{pmatrix}
        (1+|z_1|^2)\chi  
        & z_1 \bar{z}_2\chi 
        & -3\bar{x}z_1 \chi\\[0.75em]
        \bar{z_1}z_2 \chi
        &  (1+|z_2|^2)\chi  
        & -3\bar{x}z_2\chi\\[0.75em]
        -3x\bar{z}_1\chi
        & -3x\bar{z}_2 \chi
        & 9|x|^2(|z_1|^2+|z_2|^2)+\dfrac{9(r+X)}{\chi^3}
    \end{pmatrix}
\end{align} 
and
\begin{equation}
    \rho_F(|z_1|,|z_2|,|x|)=-\frac{1}{2}{\rm tr}F_{mn}F^{mn}= -\frac{72r^{4/3}}{(r+X)^2},
\end{equation}
\begin{equation}
    \rho_R(|z_1|,|z_2|,|x|)=\frac{1}{2}R_{mnpq}R^{mnpq}=\frac{120r^2}{(r+X)^{8/3}}.
\end{equation}
Note that the source is no longer proportional to $\alpha'$ as $h$ is not dependent on $\alpha'$ unlike $\Omega^2$.
In a similar fashion to the resolution of a $\mathbb{C}^2/\mathbb{Z}_2$ singularity, the delta functions are defined by employing an identity on ${\cal M}^2$ Eq.~\eqref{eq:identity}, corresponding to ${\cal M}^3$ with $z_1=0$, 
\begin{align}
    \delta_R(|z_1|,|z_2|,|x|)dz_2 \wedge d\bar{z}_2\wedge d\bar{x} \wedge dx &=-\frac{5}{12\pi^2}\frac{r^{5/3}}{(r+X)^{8/3}}dz_2 \wedge d\bar{z}_2\wedge d\bar{x} \wedge dx, \nonumber\\
    \delta_F(|z_1|,|z_2|,|x|)d\bar{z}_2\wedge dz_2 \wedge d\bar{x} \wedge dx &=-\frac{1}{4\pi^2}\frac{r}{(r+X)^{2}}dz_2 \wedge d\bar{z}_2 \wedge d\bar{x} \wedge dx,
\end{align}
and the sources can be rewritten in terms of these delta functions,
\begin{align}
    \rho_F(|z_1|,|z_2|,|x|)+
    \rho_R(|z_1|,|z_2|,|x|)=-288\pi^2 r^{1/3}\bigl[\delta_R(|z_1|,|z_2|,|x|)-\delta_F(|z_1|,|z_2|,|x|) ]. 
    \end{align}
Since the integral of these source terms over 4-cycles on ${\cal M}^3$ vanishes, it is consistent with the anomaly cancellation condition Eq.~\eqref{eq:intcondc2}. 
Both the background curvature $R$ and gauge flux $F$ are localised around the resolution of a fixed point and diverge at the origin in the limit $r, |x|\rightarrow 0$, 
where $|x|=0$ corresponds to a $\mathbb{CP}^2$. 
Such localised sources will deform the background geometry through Eq.~\eqref{eq:bianchi3}. 

To analyse the deformation of the background geometry, we rewrite the equation of motion in terms of polar coordinates for $z_{1,2}$ and $x$ in a manner similar to Section~4,
\begin{align}
    z_1=|z_1|e^{i\theta_{z_1}},\qquad
    z_2=|z_2|e^{i\theta_{z_2}},\qquad
    x=|x|e^{i\theta_{x}}.
\end{align}
The resulting metric of ${\cal M}^3$ is independent of the angular variables $\theta_{z_1},\theta_{z_2},\theta_{x}$, 
respecting the U(1) symmetries of the base $\mathbb{CP}^2$ and the fibre $\mathbb{C}$ analogous to the symmetries seen in $\mathcal{M}^2$. 
Since the source terms are also independent of these angular variables, 
we are again allowed to neglect them. 
The equation of motion is then reduced to 
\begin{multline} 
	    \frac{\chi}{(r+X)^{1/3}}\bigg[ (1+|z_1|^2) \left(\partial_{|z_1|}^2 +\frac{1}{|z_1|} \partial_{|z_1|}\right)+2|z_1||z_2| \partial_{|z_1|} \partial_{|z_2|} +(1+|z_2|^2)\left(\partial_{|z_2|}^2 +\frac{1}{|z_2|} \partial_{|z_2|}\right) \\
	    -6|x|\left( |z_1|\partial_{|z_1|} \partial_{|x|} +|z_2|\partial_{|z_2|}\partial_{|x|}\right)
	    +9\frac{|x|^2(|z_1|^2+|z_2|^2)\chi^3+(r+X)}{\chi^4}\left(\partial_{|x|}^2 +\frac{1}{|x|} \partial_{|x|}\right) \bigg] h (|z_1|,|z_2|,|x|) \\
    =-288 \pi^2 r^{1/3} [\delta_{R} (|z_1|,|z_2|,|x|) - \delta_{F} (|z_1|,|z_2|,|x|)],
\end{multline}
which is still difficult to solve analytically. Following Section 4, we will again use Mathematica to obtain numerical solutions of $h$ for different values of the blow-up parameter $r$. The boundary conditions for $|z_{1,2}|$ and $|x|$ are generalisations of those taken for the analysis on $\mathcal{M}^2$, with $\mathbb{CP}^2$ having the following boundary conditions
\begin{equation} \label{eq:zBC2}
    \left. \frac{\partial h}{\partial |z_1|} \right|_{|z_1|=0} = \left. \frac{\partial h}{\partial |z_2|} \right|_{|z_2|=0} = \left. \frac{\partial h}{\partial |z_1|} \right|_{|z_1|= \infty} = \left. \frac{\partial h}{\partial |z_2|} \right|_{|z_2|= \infty} = 0,
\end{equation}
and the complex line over $\mathbb{CP}^2$ having the following boundary conditions
\begin{align} \label{eq:xBC2}
    \left. \frac{\partial h}{\partial |x|} \right|_{|x|=0} &= 0, &
    h(|z_1|,|z_2|,|x|\rightarrow\infty) &= 0,
\end{align}
where the Dirichlet boundary condition is now set to 0 to recover the background metric far away from the resolution $\mathcal{M}^3$ as we are working with the trace of the metric instead of a conformal factor.
Again, for computational purposes, we impose boundary conditions for $|z_{1,2}|$ and $|x|$ along a sufficiently large, but finite radius $L$ instead of at infinity to obtain a numerical approximation. Figures \ref{fig:6D_r=3} and \ref{fig:6D_r=1} below show plots for the numerical solution of $h$ and the trace of the background metric $g$ for blow-up parameters $r=3$ and $1$. In both cases, we have set\footnote{The reduction in $L$ here is due to limitations in computing power.} $L=5$. 

 \begin{figure}[htbp]
  \begin{minipage}{0.3\hsize}
   \begin{center}
     \includegraphics[width=50mm]{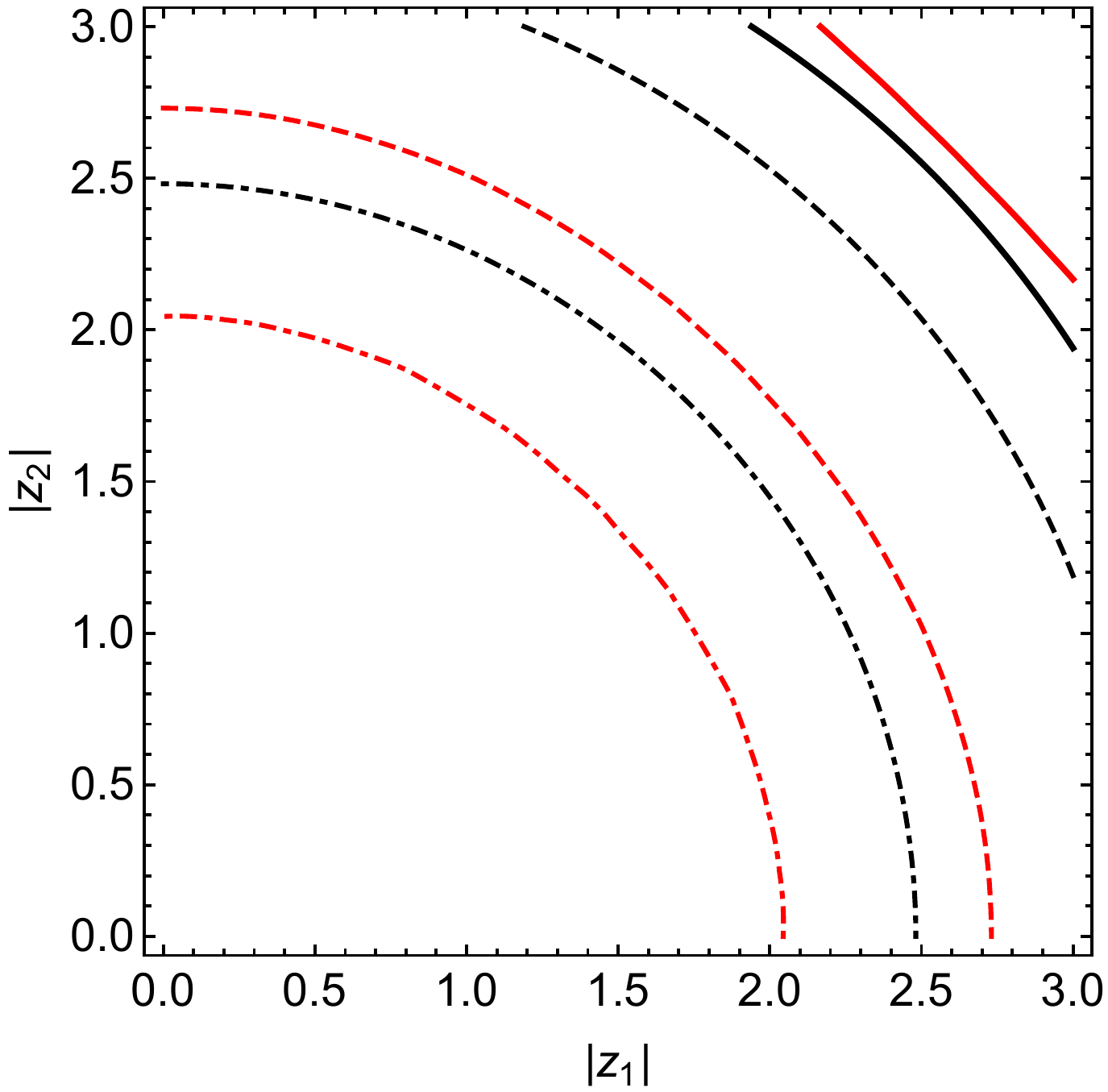}
    \end{center}
   \end{minipage}
  \begin{minipage}{0.3\hsize}
   \begin{center}
    \includegraphics[width=50mm]{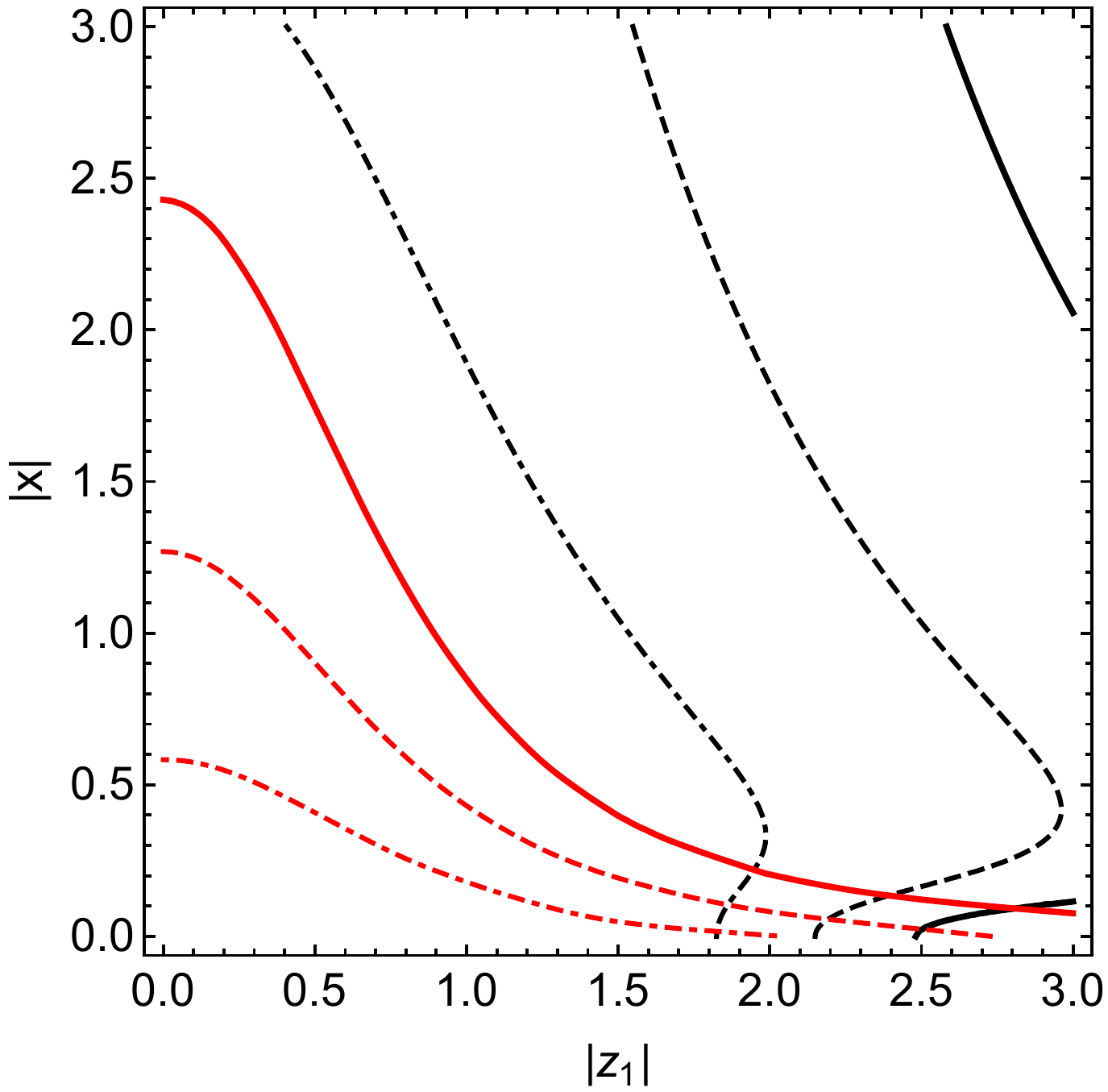}
   \end{center}
  \end{minipage}
  \begin{minipage}{0.3\hsize}
   \begin{center}
    \includegraphics[width=50mm]{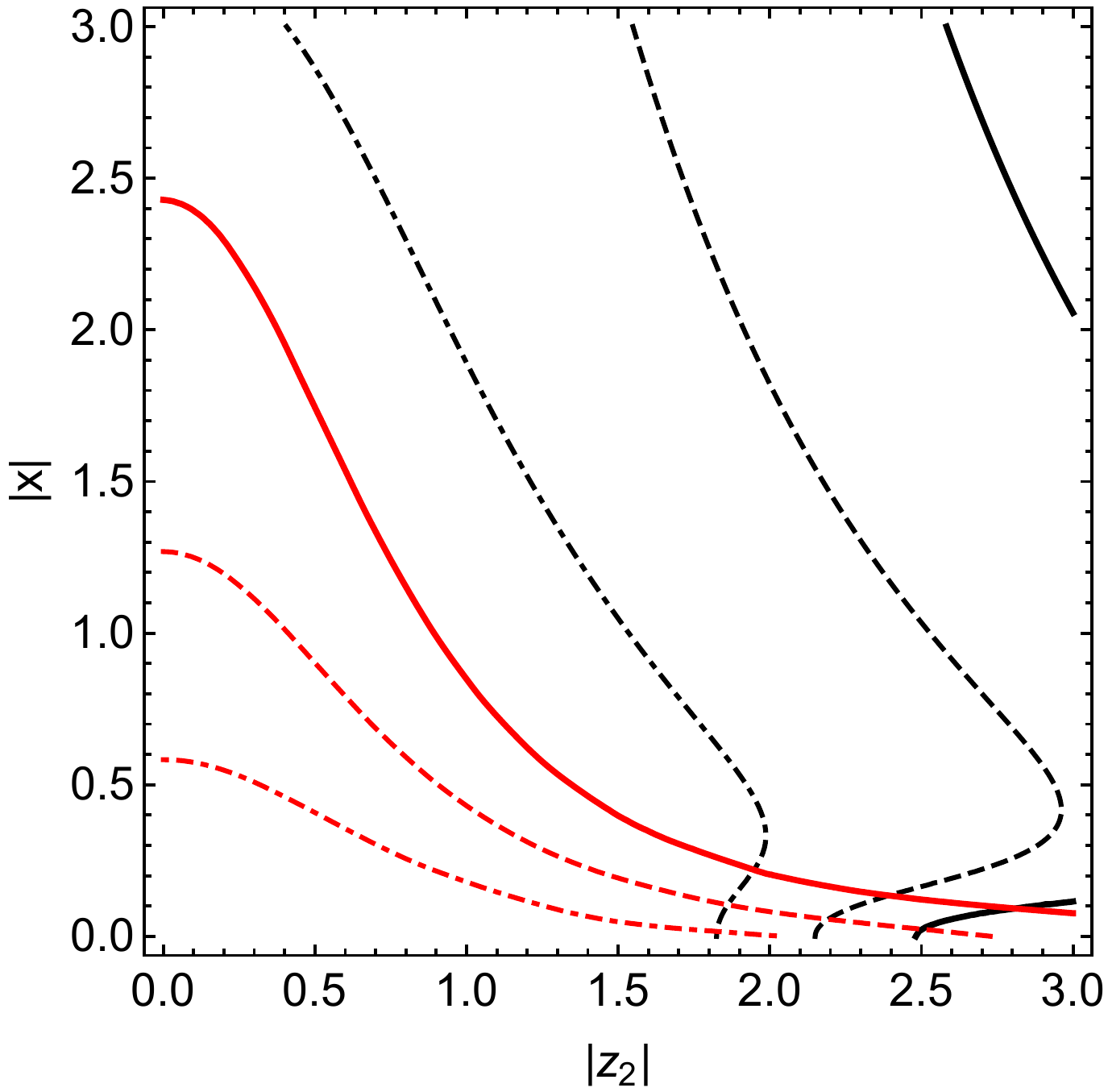}
   \end{center}
  \end{minipage}
    \caption{Numerical solution for the trace of the metric correction $h$ and background metric $g$ on the resolution $\mathcal{M}^3$ with blow-up parameter $r=3$. $h$ and $g$ are drawn in red and black curves in all panels. 
    The left panel shows the $(|z_1|, |z_2|)$-plane with $|x|=0$. Red (black) solid, dashed, dot-dashed curves denote $h = -0.1, -0.3, -0.5$, (${\rm tr}(g)= 140, 80, 20$) respectively. In the middle ($(|z_1|, |x|)$-plane with $|z_2|=0$) and right panels ($(|z_2|, |x|)$-plane with $|z_1|=0$), the red (black) solid, dashed, dot-dashed curves denote $h = -0.1, -0.3, -0.5$, (${\rm tr}(g)= 20, 10, 5$) respectively.}
    \label{fig:6D_r=3}
\end{figure}

 \begin{figure}[htbp]
  \begin{minipage}{0.3\hsize}
   \begin{center}
     \includegraphics[width=50mm]{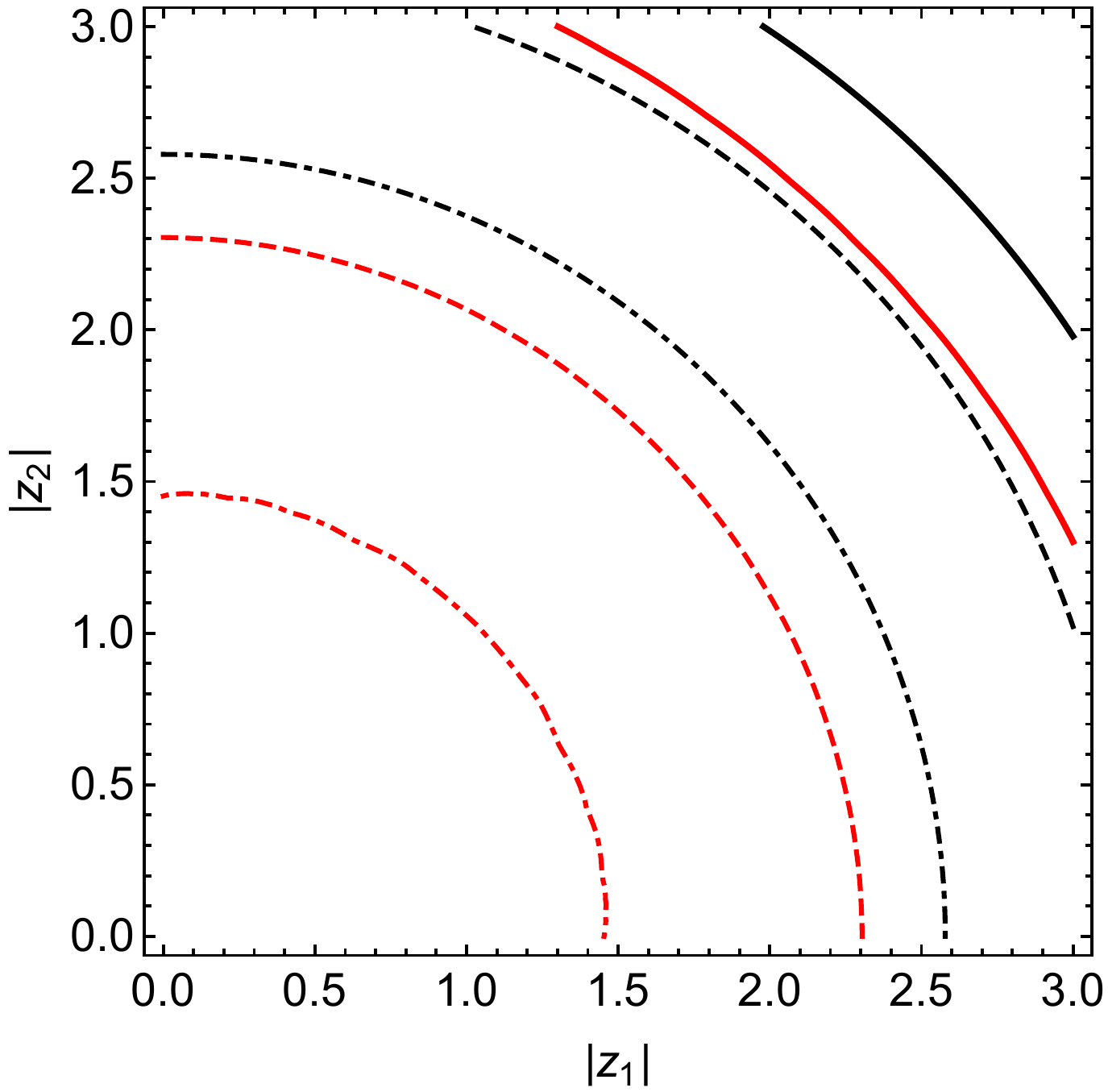}
    \end{center}
   \end{minipage}
  \begin{minipage}{0.3\hsize}
   \begin{center}
    \includegraphics[width=50mm]{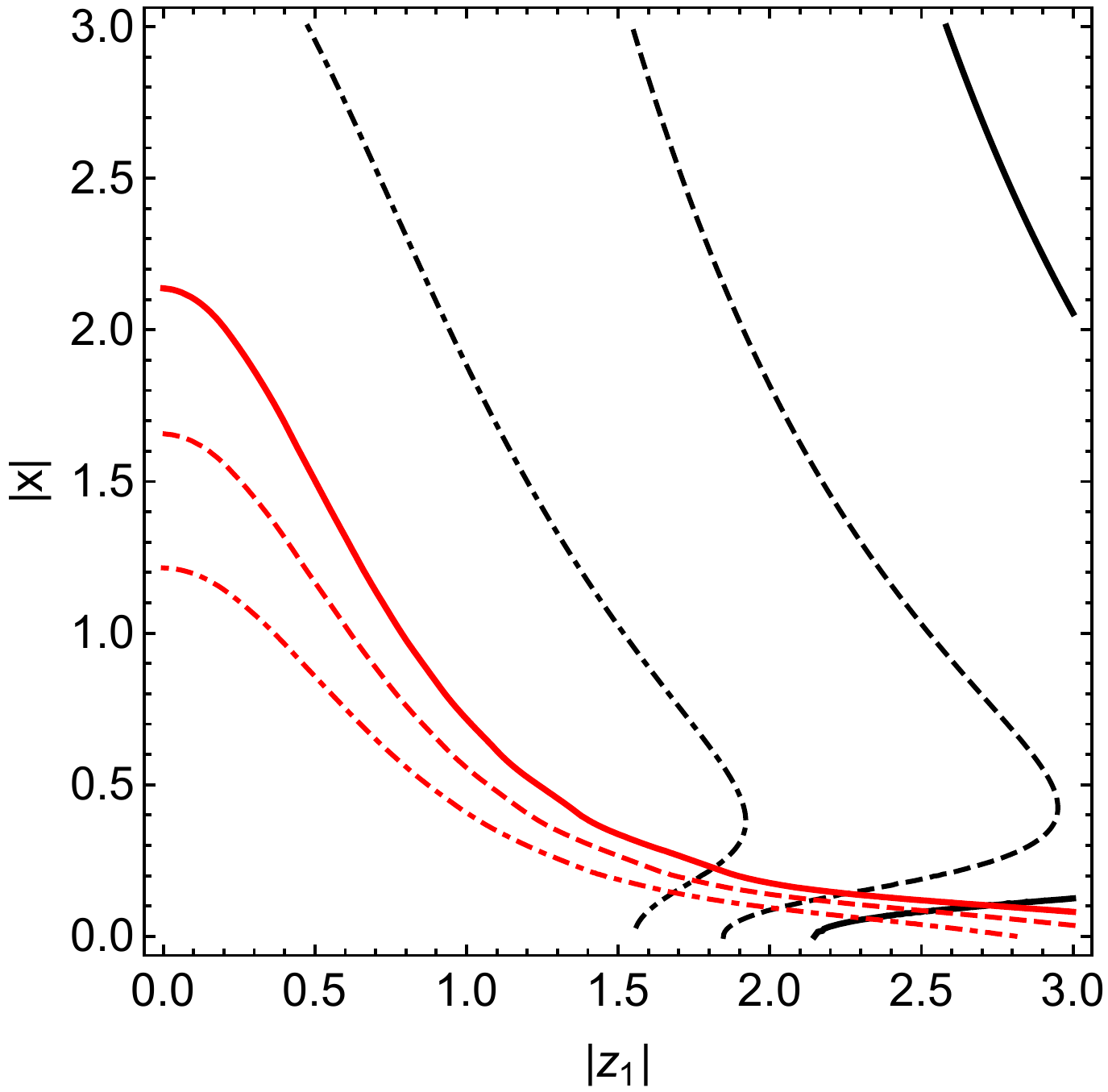}
   \end{center}
  \end{minipage}
  \begin{minipage}{0.3\hsize}
   \begin{center}
    \includegraphics[width=50mm]{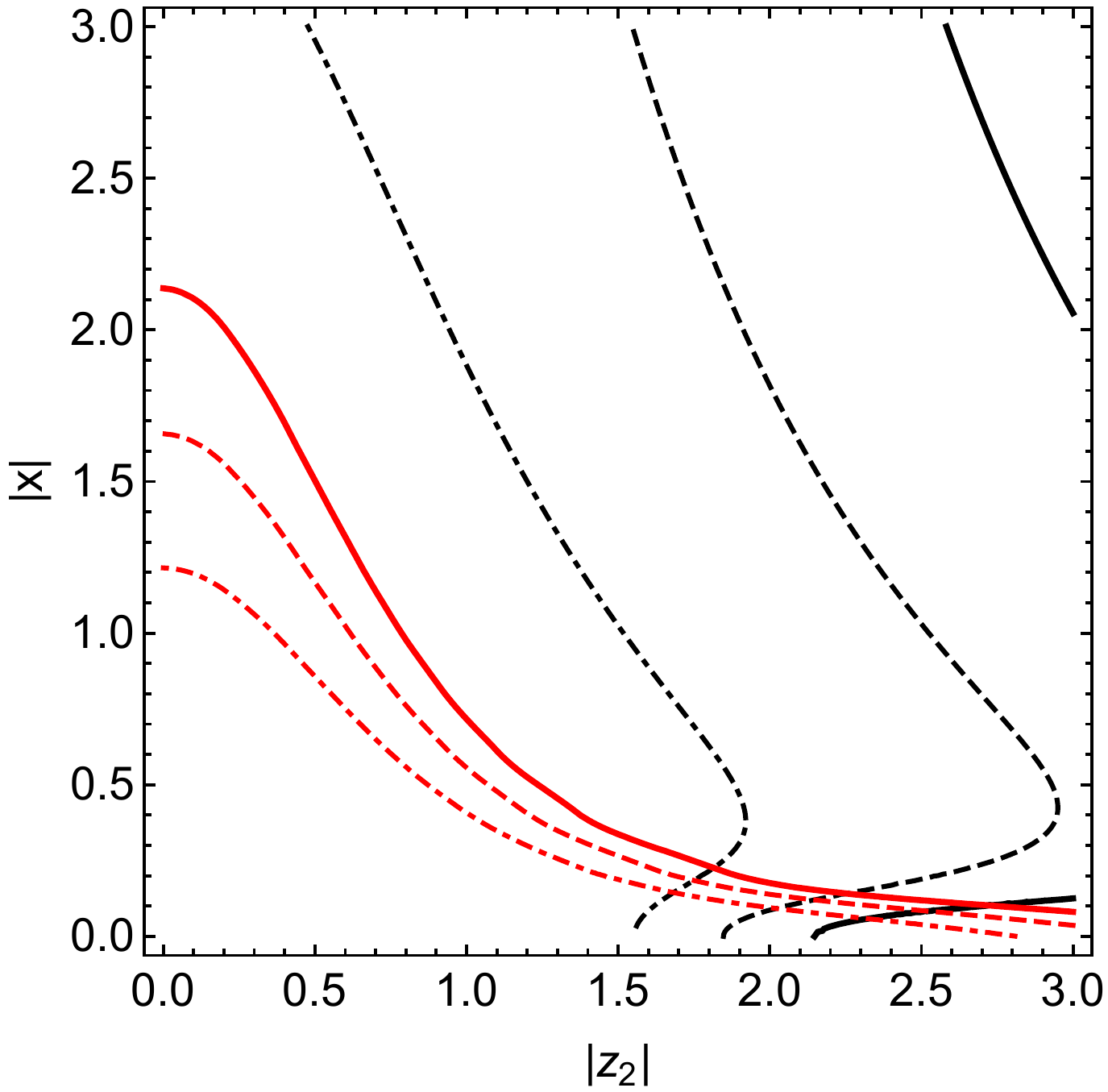}
   \end{center}
  \end{minipage}
    \caption{Numerical solution for the trace of the metric correction $h$ and background metric $g$ on the resolution $\mathcal{M}^3$ with blow-up parameter $r=1$. $h$ and $g$ are drawn in red and black curves in all panels. 
    The left panel shows the $(|z_1|, |z_2|)$-plane with $|x|=0$. Red (black) solid, dashed, dot-dashed curves denote $h = -0.1, -0.4, -0.8$, (${\rm tr}(g)= 300, 150, 50$) respectively. In the middle ($(|z_1|, |x|)$-plane with $|z_2|=0$) and right panels ($(|z_2|, |x|)$-plane with $|z_1|=0$), the red (black) solid, dashed, dot-dashed curves denote $h = -0.05, -0.1, -0.2$, (${\rm tr}(g)= 20, 10, 5$) respectively.}
    \label{fig:6D_r=1}
\end{figure}

From Figures \ref{fig:6D_r=3} and \ref{fig:6D_r=1}, we see that the behaviour of $h$ on the $(|z_{1,2}|,|x|)$-planes for $\mathcal{M}^3$ is similar to that found for $\Omega^2$ on the $(|z|,|x|)$-plane of $\mathcal{M}^2$, and that the solution $h$ is symmetrical on the $(|z_{1}|,|z_{2}|)$-plane. On closer inspection, we note that $h$ actually takes negative values as compared to the $\mathcal{M}^2$ case. This can be attributed to the change in boundary conditions (cf. Eqs.~\eqref{eq:xBC} and \eqref{eq:xBC2}) and does not mean that we have encountered negative metrics for both the 6-dimensional toroidal orbifold and its resolution. Recall that $h$ represents the trace of the metric correction, not a conformal factor that scales the background metric. This means that a negative $h$ is admissible provided that the trace of the full metric $G$ is positive. This is indeed the case as the trace of the background metric $g$ is positive and of a larger magnitude than the metric correction $h$. Like its 4-dimensional counterpart, the metric correction $h$ is localised around the origin and is small in magnitude. This suggests that at the $O(\alpha')$ level, the trace of the background metric is only deformed by a small fraction, with the majority happening near the centre of the fixed point. The same statement as $\mathcal{M}^2$ holds - non-standard embedding at $O(\alpha')$ is tractable given that the integrated Bianchi identity Eq.~\eqref{eq:M3intcond} is satisfied.

\section{Conclusion and Discussion}

In this paper, we studied $O(\alpha')$ corrections to the compact space metric of 10D heterotic supergravity. We adopted a conformal factor ansatz for orbifold limits of K3 and the resolution of a $T^4/\mathbb{Z}_2$ orbifold singularity and found that in the orbifold limit case, the conformal factor becomes negative around certain fixed points in the non-standard embedding scenario, leading to the need to consider higher order $\alpha'$-corrections and/or background torsion. This leaves us with a constant conformal factor in the standard embedding scenario as the only viable option at $O(\alpha')$. In contrast, the resolution $\mathcal{M}^2$ allows for non-standard embedding without NS5 branes as the anomaly cancellation condition can be satisfied without generating a negative conformal factor. This means that we can obtain a non-constant conformal factor at $O(\alpha')$ for the case of $\mathcal{M}^2$ being the background geometry. We note that the full metric in this case corresponds exactly to the shift in the K\"ahler form required to stabilise K\"ahler moduli. The numerical solution to the conformal factor on $\mathcal{M}^2$ also shows that the effects of the deformation are rather localised around the centre of the fixed point. This is attributed to the localised nature of the curvature and flux which sources the deformation. We have also generalised these results to the $T^6$ orbifold by considering the trace of the metric correction instead of a conformal factor as the ansatz is known to not hold in dimensions other than 4. We obtained results that behave analogously to the 4-dimensional case, and similarly saw that the deformation to the trace of the background metric at $O(\alpha')$ is small and mostly localised towards the origin. We cannot use this result to deduce the behaviour of the separate components of the metric correction as $h$ corresponds to a part of the corrections to the metric.

There are several implications that can be garnered from the results in this paper. First of all, it would be interesting to consider deformations to not just the geometry, but the CFTs of toroidal orbifold models or their resolutions, where the CFTs can be calculated exactly. It would also be instructive to study the stabilisation of K\"ahler moduli as it is possible to explicitly construct the potential for these moduli. In particular, this entertains the possibility of applying the $\mathcal{M}^2$ geometry in moduli stabilisation and exploring scenarios similar to KKLT \cite{Kachru:2003aw}. We can also apply these results to the generation of 4D effective Yukawa couplings and perhaps reconstruct realistic or semi-realistic Standard Model parameters. It is known that the presence of fluxes on internal manifolds can act to localise wavefunctions around such sources. In the context of the K3 geometries considered in this paper, this would suggest that for the orbifold limit of K3, the wavefunctions are unaffected at $O(\alpha')$ for the standard embedding scenario, whilst for the resolution $\mathcal{M}^2$ the wavefunctions could localise around the centre due to the non-trivial conformal factor in the non-standard embedding scenario. It will be interesting to investigate how these zero modes are shifted or localised compared to their uncorrected counterparts. By calculating these wavefunctions using well-known techniques (see \cite{Cre04,Conlon:2008qi}) it should be possible in principle to determine the Yukawa coupling via their triple intersection integrals, and compare how the $O(\alpha')$ correction affects them. This allows for a more diverse class of geometries for string model building and is a novel approach to the generation of certain Standard Model parameters. The negative metric problem from the non-standard embedding scenario of $T^4/\mathbb{Z}_n$ remains to be explored, and the inclusion of a background torsion in our analysis should provide some answers.

\section*{Acknowledgments}
  We would like to thank H.~Abe for useful discussions. P. L. would like to thank M. Bowen and M. Honda for fruitful discussions.
  H.~O. was supported in part by Grant-in-Aid for Young Scientists (B) (No.~17K14303) 
  from the Japan Society for the Promotion of Science. P. L. was supported in part by the SEI Group CSR Foundation Scholarship.

\appendix
\section{Background Curvature and U(1) Flux on Resolutions of $\mathbb{C}^n/\mathbb{Z}_n$ Orbifolds}
\label{app}

Here, we will briefly summarise the construction of background curvature and U(1) flux 2-forms on resolutions of $\mathbb{C}^n/\mathbb{Z}_n$ orbifolds, ${\cal M}^n$ along the lines of \cite{Nib07}. As mentioned previously, $\mathcal{M}^n$ is constructed as the complex line bundle over $\mathbb{CP}^{n-1}$. Let $x$ be the coordinate that parametrises the complex line over $\mathbb{CP}^{n-1}$, and $z$ be the vector that contains the local coordinates of $\mathbb{CP}^{n-1}$
\begin{equation}
    z = \begin{pmatrix}
        z_1 \\
        \vdots \\
        z_{n-1}
    \end{pmatrix}.
\end{equation}
The K\"ahler potential reproducing the $\mathbb{C}^n/\mathbb{Z}_n$ orbifold in the blow down limit is given by
\begin{align}
    K(X)=\frac{1}{n}\int_1^X \frac{dX'}{X'}(r+X')^{\frac{1}{n}},
\end{align}
as a function of $X=\bar{x}\chi^n x$ with $\chi=1+\bar{z}z$, from which the K\"ahler metric is obtained as in Eq.~\eqref{eq:resmetric} for $n=2$ and Eq.~\eqref{eq:resmetricT6} for $n=3$. 
For the sake of defining the gauge bundle on this resolution space, 
we introduce a vector $e$ of $(n-1)$ 1-forms containing the holomorphic vielbeins of $\mathbb{CP}^{n-1}$ and the 1-form $\epsilon$ associated with a complex line bundle,
\begin{align}
    e=\chi^{-\frac{1}{2}}\tilde{\chi}^{-\frac{1}{2}}dz,\qquad
    \epsilon=dy +ni{\cal B}y,
\end{align}
written in terms of a convenient coordinate $y=\chi^{\frac{n}{2}}x$ parametrising the complex line over $\mathbb{CP}^{n-1}$. 
Here, $\tilde{\chi}$ is an $n-1 \times n-1$ matrix $\mathbbm{1}_{n-1} + z\bar{z}$ and $i{\cal B}$ denotes a U(1) connection originating from the trace of the U$(n-1)$ 1-form $i\tilde{{\cal B}}$ on $\mathbb{CP}^{n-1}$. 
\begin{equation}
    i\mathcal{B}=-\mathrm{tr}(i\tilde{\mathcal{B}}) = \frac{1}{2}(\bar{z}e-\bar{e}z),
\end{equation}
\begin{equation}
    i\tilde{\mathcal{B}}=\tilde{\chi}^{-\frac{1}{2}}\bar{\partial}(\tilde{\chi}^{\frac{1}{2}}) - \partial(\tilde{\chi}^{\frac{1}{2}})\tilde{\chi}^{-\frac{1}{2}}
\end{equation}
The K\"ahler metric is then expressed as 
\begin{equation} \label{eq:kahlermetric}
    g=\bar{E}\otimes E,
\end{equation} 
where
\begin{align}
E=
\begin{pmatrix}
    \sqrt{(r+X)^{\frac{1}{n}}}\,e
    \\[0.5em]
    \dfrac{1}{n}\sqrt{(r+X)^{\frac{1}{n}-1}}\,\epsilon
\end{pmatrix}
\end{align}
is the holomorphic vielbein 1-form. In component form, Eq.~\eqref{eq:kahlermetric} reads
\begin{equation}
    g_{\bar{i}j} = \bar{E}_{\bar{i}}^{\bar{a}} E_j^b \delta_{\bar{a}b}.
\end{equation}
This statement indicates the fact that our manifold is locally K\"ahler.

By solving the Cartan structure equation $dE+ \omega \wedge E=0$, the spin connection $\omega$ can be obtained as
\begin{align}
    \omega=
    \begin{pmatrix}
    i\tilde{{\cal B}}-{i\cal B}
    +\dfrac{1}{2n}\dfrac{\bar{y}\epsilon-\bar{\epsilon} y}{r+X} 
    & \dfrac{\bar{y} \epsilon}{\sqrt{r+X}}
    \\[0.75em]
    -\dfrac{\bar{e} y}{\sqrt{r+X}}
    &
    ni{\cal B}-\dfrac{n-1}{2n}\dfrac{\bar{y} \epsilon-\bar{\epsilon} y}{r+X} 
    \end{pmatrix}
    .
\end{align}
From this expression, the curvature 2-form $R$ can be calculated using $R=d\omega +\omega \wedge \omega$ to give
\begin{align}
    R = \frac{r}{r+X}
    \begin{pmatrix}
    e \wedge\bar{e}-\bar{e} \wedge e +\dfrac{1}{n}\dfrac{\bar{\epsilon}\wedge \epsilon}{r+X} 
    & \dfrac{\bar{\epsilon}\wedge e}{\sqrt{r+X}}
    \\[0.75em]
    \dfrac{\bar{e}\wedge \epsilon}{\sqrt{r+X}}
    &
    n\bar{e}\wedge e-\dfrac{n-1}{n}\dfrac{\bar{\epsilon}\wedge \epsilon}{r+X} 
    \end{pmatrix}
    ,
\end{align}
which is traceless, reflecting the SU($n$) holonomy of ${\cal M}^n$. 

In addition, the U(1) gauge connection and supersymmetric flux can be constructed as
\begin{align}
    iA=i{\cal B}+\frac{1}{2nX}\biggl[1-\left(\frac{r}{r+X}\right)^{1-\frac{1}{n}}\biggl](\bar{\epsilon}y-\bar{y}\epsilon),
\end{align}
and
\begin{align}
    iF=\left(\frac{r}{r+X}\right)^{1-\frac{1}{n}}\left(\bar{e}\wedge e-\frac{n-1}{n^2}\frac{\bar{\epsilon}\wedge \epsilon}{r+X}\right).
\end{align}
The embedding of U(1)s into SO(32) or $E_8 \times E_8$ is defined by
\begin{equation}
    iF_V = iF V^I H_I,
\end{equation}
where $V^I$ and $H_I$ are the shift vector and Cartan generator of the subgroup corresponding to the $I$th U(1) for $I=1,... ,16$. The U(1) flux $iF_V$ is quantized on the $\mathbb{CP}^1$ at $x=0$ on ${\cal M}^n$. The normalisation adopted for the trace of the Cartan generators is
\begin{equation}
    \mathrm{tr}(H_I H_J) = \delta_{IJ}.
\end{equation}
The traces in the SO(32) and $E_8 \times E_8$ gauge groups are related by
\begin{equation}
    \mathrm{tr}(iF_V \wedge iF_V) = \frac{1}{30}\mathrm{Tr}(iF_V \wedge iF_V),
\end{equation}
where $\mathrm{tr}$ denotes the trace in the fundamental representation of SO(32) and $\mathrm{Tr}$ denotes the trace in the adjoint representation of SO(32) and $E_8 \times E_8$ \cite{Blumenhagen:2013fgp}.

\section{Symmetry of Differential Equations on $\mathcal{M}^2$}
\label{app2}

Using a bit of rigour, one can show that the differential equation Eq.~\eqref{eq:bigpoisson} also inherits the symmetries of $\mathcal{M}^2$. We can express the Laplacian in polar coordinates using Eq.~\eqref{eq:polarresmetric}, and rewrite Eq.~\eqref{eq:bigpoisson} as
\begin{multline} \label{eq:fullpolarpoisson}
	    \bigg[ \frac{\chi^2}{\sqrt{r+X}} \left(\partial_{|z|}^2+\frac{1}{|z|}\partial_{|z|}+\frac{1}{|z|^2} \partial_{\theta_z}^2\right) - \frac{4\chi}{\sqrt{r+X}} \left(|z||x|\partial_{|z|} \partial_{|x|}+\partial_{\theta_z}\partial_{\theta_x}\right) \\[0.5em]
	    + \left( \frac{4 \sqrt{r+X}}{\chi^2} + \frac{4|x|^2 |z|^2}{\sqrt{r+X}}\right) \left(\partial_{|x|}^2+\frac{1}{|x|}\partial_{|x|}+\frac{1}{|x|^2}\partial_{\theta_x}^2\right) \bigg] \Omega^2
    = \frac{\alpha'}{8}\left[-\frac{12r}{(r+X)^2} + \frac{24r^2}{(r+X)^3}\right].
\end{multline}
A partial differential equation has a symmetry group $\mathfrak{G}$ if, given a solution $u(x)$, for any element $\mathfrak{g} \in \mathfrak{G}$, the transformed solution $\tilde{u}(\tilde{x}) = \mathfrak{g} \cdot u(x)$ is also a solution \cite{Olv92}. To demonstrate this for Eq.~\eqref{eq:fullpolarpoisson}, we will take $\mathfrak{G}=U(1)_{\mathbb{CP}^1} \times U(1)_{\mathbb{C}}$ and let $\mathfrak{g}$ act on the coordinates as
\begin{align} \label{eq:frakgtrans}
    \mathfrak{g} : (|z|,\theta_z,|x|,\theta_x) &\rightarrow (|z|',\theta_z',|x|',\theta_x') = (|z|, \theta_z + \varepsilon, |x|, \theta_x + \eta), & \varepsilon,\eta &\in \mathbb{R}.
\end{align}
The transformed solution is then
\begin{equation}
    \Omega'^2(|z|',\theta_z',|x|',\theta_x') = \mathfrak{g}\cdot\Omega^2(|z|,\theta_z,|x|,\theta_x).
\end{equation}
To see whether this is a solution to the Poisson equation Eq.~\eqref{eq:fullpolarpoisson}, simply apply the Laplace operator on the left hand side of Eq.~\eqref{eq:fullpolarpoisson} to the transformed solution above. In compact notation, we have
\begin{equation}
\begin{split}
    \nabla^2 [\mathfrak{g}\cdot\Omega^2] &= \nabla'^2 \Omega'^2 \\
    &= \mathfrak{g}\cdot[\nabla^2 \Omega^2] \\
    &= \mathfrak{g} \cdot \rho(|z|,|x|) \\
    &= \rho(|z|,|x|).
\end{split}
\end{equation}
The argument is as follows. By straightforward application of the chain rule, partial derivatives appearing in the rotated coordinates can be shown to be equivalent to those before rotation, i.e., $\partial'_m = \partial_m$ for $m=|z|,\theta_z,|x|,\theta_x$. This is easily understood from the fact that the transformation parameters $\varepsilon$ and $\eta$ are real constants. Additionally, the coefficients of the Laplacian are unaffected by the shift due to $\mathfrak{g}$ since they do not depend on $\theta_z$ and $\theta_x$. This leaves us with a Laplacian that is invariant under rotations, which is to be expected given the metric itself has the same invariance. This allows us to write the first equality. The second equality then comes from the fact that we can view $\nabla'^2 \Omega'^2$ as the whole left hand side of Eq.~\eqref{eq:fullpolarpoisson} being rotated by $\mathfrak{g}$. Using Poisson's equation, we can rewrite $\nabla^2 \Omega^2$ as the source $\rho$, giving us the third line. Since the source is rotationally invariant, the action of $\mathfrak{g}$ on it is trivial. The crux of the argument then lies in the fact that our source is rotationally invariant, and this coupled with a rotationally invariant geometry tells us that our differential equation Eq.~\eqref{eq:fullpolarpoisson} possesses the same $U(1)_{\mathbb{CP}^1} \times U(1)_{\mathbb{C}}$ symmetry. Accordingly, it is reasonable to find solutions that respect the same symmetry in the underlying system, and are independent of the angular variables $\theta_z$ and $\theta_x$. This allows us to neglect any terms in the equation of motion Eq.~\eqref{eq:fullpolarpoisson} containing $\partial_{\theta_z}$ or $\partial_{\theta_x}$. The result is Eq.~\eqref{eq:polarpoisson}, the reduced form of the equation of motion for the conformal factor on $\mathcal{M}^2$. 

More technically, in general, although symmetries can be used to reduce the number of independent variables of a differential equation, the form of the reduced equation depends on group invariants, rather than simply ignoring certain variables. The solution to the reduced equations are then functions of these group invariants. In our case, the two-parameter transformation group $\mathfrak{G}$ lets us get rid of two independent variables. The group invariants are then any function of $|z|$ and $|x|$ as they are unaffected by $\mathfrak{g}$. This allows us to ignore the angular variables in our reduced equation. For more details, see \cite{Olv92,Olv86}.

\bibliography{refs}        

\providecommand{\href}[2]{#2}\begingroup\raggedright\begin{thebibliography}{10}

\bibitem{Balasubramanian:2005zx}
V.~Balasubramanian, P.~Berglund, J.~P. Conlon, and F.~Quevedo, {\it
  {Systematics of moduli stabilisation in Calabi-Yau flux compactifications}},
  {\em JHEP} {\bf 03} (2005) 007,
  [\href{http://xxx.lanl.gov/abs/hep-th/0502058}{{\tt hep-th/0502058}}].

\bibitem{Becker:2002nn}
K.~Becker, M.~Becker, M.~Haack, and J.~Louis, {\it {Supersymmetry breaking and
  alpha-prime corrections to flux induced potentials}},  {\em JHEP} {\bf 06}
  (2002) 060, [\href{http://xxx.lanl.gov/abs/hep-th/0204254}{{\tt
  hep-th/0204254}}].

\bibitem{Ang10}
{L. Anguelova, C. Quigley, and S. Sethi}, {\it {The leading quantum corrections
  to stringy K{\"a}hler potentials}},  {\em {JHEP}} {\bf {2010}} ({2010}) {65}.

\bibitem{Ibanez:1987sn}
L.~E. Ibanez, J.~E. Kim, H.~P. Nilles, and F.~Quevedo, {\it {Orbifold
  Compactifications with Three Families of $SU(3) \times SU(2) \times
  U(1)^n$}},  {\em Phys. Lett.} {\bf B191} (1987) 282--286.

\bibitem{Bailin:1987xm}
D.~Bailin, A.~Love, and S.~Thomas, {\it {A Three Generation Orbifold
  Compactified Superstring Model With Realistic Gauge Group}},  {\em Phys.
  Lett.} {\bf B194} (1987) 385--389.

\bibitem{Ibanez:1987pj}
L.~E. Ibanez, J.~Mas, H.-P. Nilles, and F.~Quevedo, {\it {Heterotic Strings in
  Symmetric and Asymmetric Orbifold Backgrounds}},  {\em Nucl. Phys.} {\bf
  B301} (1988) 157--196.

\bibitem{Casas:1988hb}
J.~A. Casas and C.~Munoz, {\it {Three Generation $SU(3) \times SU(2) \times
  U(1)_Y$ Models from Orbifolds}},  {\em Phys. Lett.} {\bf B214} (1988) 63--69.

\bibitem{Font:1989aj}
A.~Font, L.~E. Ibanez, F.~Quevedo, and A.~Sierra, {\it {The Construction of
  \enquote*{Realistic} Four-Dimensional Strings Through Orbifolds}},  {\em
  Nucl. Phys.} {\bf B331} (1990) 421--474.

\bibitem{Hamidi:1986vh}
S.~Hamidi and C.~Vafa, {\it {Interactions on Orbifolds}},  {\em Nucl. Phys.}
  {\bf B279} (1987) 465--513.

\bibitem{Aspinwall:1994ev}
P.~S. Aspinwall, {\it {Resolution of orbifold singularities in string theory}},
   \href{http://xxx.lanl.gov/abs/hep-th/9403123}{{\tt hep-th/9403123}}. [AMS/IP
  Stud. Adv. Math.1,355(1996)].

\bibitem{Nib07}
{S. G. Nibbelink, M. Trapletti, and M. G. A. Walter}, {\it {Resolutions of
  $\mathbb{C}^n/\mathbb{Z}_n$ Orbifolds, their $U(1)$ Bundles, and Applications
  to String Model Building}},  {\em {JHEP}} {\bf {2007}} ({2007}) {035}.

\bibitem{Nib08}
{S. G. Nibbelink, T. Ha, M. Trapletti}, {\it {Toric resolutions of heterotic
  orbifolds}},  {\em {Phys. Rev.}} {\bf {D 77}} ({2008}) {19}.

\bibitem{Dix85}
{L. Dixon, J. A. Harvey, C. Vafa, and E. Witten}, {\it {Strings on orbifolds}},
   {\em {Nucl. Phys.}} {\bf {B261}} ({1985}) {678--686}.

\bibitem{Dixon:1986jc}
L.~J. Dixon, J.~A. Harvey, C.~Vafa, and E.~Witten, {\it {Strings on orbifolds
  (II)}},  {\em Nucl. Phys.} {\bf B274} (1986) 285--314.

\bibitem{Kat90}
{Y. Katsuki, Y. Kawamura, T. Kobayashi, Y. Ono, and K. Tanioka}, {\it {$Z_N$
  Orbifold Models}},  {\em {Nucl. Phys.}} {\bf {B341}} ({1990}) {611--640}.

\bibitem{Bailin:1999nk}
D.~Bailin and A.~Love, {\it {Orbifold compactifications of string theory}},
  {\em Phys. Rept.} {\bf 315} (1999) 285--408.

\bibitem{Ibanez:2012zz}
L.~E. Ibanez and A.~M. Uranga, {\em {String theory and particle physics: An
  introduction to string phenomenology}}.
\newblock Cambridge University Press, 2012.

\bibitem{Blumenhagen:2013fgp}
R.~Blumenhagen, D.~Lüst, and S.~Theisen, {\em {Basic concepts of string
  theory}}.
\newblock Theoretical and Mathematical Physics. Springer, Heidelberg, Germany,
  2013.

\bibitem{Str86}
{A. Strominger}, {\it {Superstrings with torsion}},  {\em {Nucl. Phys.}} {\bf
  {B 274}} ({1986}) {253}.

\bibitem{Candelas:1985en}
P.~Candelas, G.~T. Horowitz, A.~Strominger, and E.~Witten, {\it {Vacuum
  Configurations for Superstrings}},  {\em Nucl. Phys.} {\bf B258} (1985)
  46--74.

\bibitem{Green:1987mn}
M.~B. Green, J.~H. Schwarz, and E.~Witten, {\em {Superstring Theory: Volume 2,
  Loop Amplitudes, Anomalies and Phenomenology}}.
\newblock Cambridge Monographs on Mathematical Physics. Cambridge University
  Press, 1988.

\bibitem{Green:1984sg}
M.~B. Green and J.~H. Schwarz, {\it {Anomaly Cancellations in Supersymmetric
  D=10 Gauge Theory and Superstring Theory}},  {\em Phys. Lett.} {\bf 149B}
  (1984) 117--122.

\bibitem{Bec06}
{K. Becker, M. Becker, and J. Schwarz}, {\em {String Theory and M-Theory: A
  Modern Introduction}}.
\newblock {Cambridge University Press}, {New York}, {2006}.

\bibitem{Gil03}
{J. Gillard, G. Papadopoulos, and D. Tsimpis}, {\it {Anomaly, Fluxes and (2,0)
  Heterotic-String Compactifications}},  {\em {JHEP}} {\bf {2003}} ({2003})
  {035}.

\bibitem{Car04}
{S. Carroll}, {\em {Spacetime and Geometry: An Introduction to General
  Relativity}}.
\newblock {Addison-Wesley}, {San Francisco}, {2004}.

\bibitem{Wal88}
{M. A. Walton}, {\it {Heterotic string on the simplest Calabi-Yau manifold and
  its orbifold limits}},  {\em {Phys. Rev.}} {\bf {D 37}} ({1988}) {377}.

\bibitem{Mam14}
{M. Mamode}, {\it {Fundamental solution of the Laplacian on flat tori and
  boundary value problems for the planar Poisson equation in rectangles}},
  {\em {Bound. Val. Probl.}} {\bf {2014}} ({2014}) {221}.

\bibitem{Pol98}
{J. Polchinski}, {\em {String Theory, volume I: An Introduction to the Bosonic
  String}}.
\newblock {Cambridge University Press}, {New York}, {1998}.

\bibitem{Shi00}
{B. R. Greene, K. Schalm, and G. Shiu}, {\it {Warped compactifications in M and
  F theory}},  {\em {Nucl. Phys.}} {\bf {B 584}} ({2000}) {480}.

\bibitem{Carlevaro:2013vla}
L.~Carlevaro and S.~Groot~Nibbelink, {\it {Heterotic warped Eguchi-Hanson
  spectra with five-branes and line bundles}},  {\em JHEP} {\bf 10} (2013) 097,
  [\href{http://xxx.lanl.gov/abs/1308.0515}{{\tt arXiv:1308.0515}}].

\bibitem{Egu78}
{T. Eguchi and A. J. Hanson}, {\it {Asymptotically flat self-dual solutions to
  euclidean gravity}},  {\em {Phys. Lett.}} {\bf {B 74}} ({1978}) {249--251}.

\bibitem{Egu80}
{T. Eguchi, P. B. Gilkey, and A. J. Hanson}, {\it {Gravitation, gauge theories
  and differential geometry}},  {\em {Phys. Rep.}} {\bf {66}} ({1980})
  {213--393}.

\bibitem{Calabi}
{E. Calabi}, {\it {Métriques kählériennes et fibrés holomorphes}},  {\em
  {Ann. Scient. Ec. Norm. Sup.}} {\bf {12}} ({1979}) {269}.

\bibitem{Kachru:2003aw}
S.~Kachru, R.~Kallosh, A.~D. Linde, and S.~P. Trivedi, {\it {De Sitter vacua in
  string theory}},  {\em Phys. Rev.} {\bf D68} (2003) 046005,
  [\href{http://xxx.lanl.gov/abs/hep-th/0301240}{{\tt hep-th/0301240}}].

\bibitem{Cre04}
{D. Cremades, L.E. Ibá{\~n}ez, and F. Marchesano}, {\it {Computing Yukawa
  Couplings from Magnetized Extra Dimensions}},  {\em {JHEP}} {\bf {2004}}
  ({2004}) {079}.

\bibitem{Conlon:2008qi}
J.~P. Conlon, A.~Maharana, and F.~Quevedo, {\it {Wave Functions and Yukawa
  Couplings in Local String Compactifications}},  {\em JHEP} {\bf 09} (2008)
  104, [\href{http://xxx.lanl.gov/abs/0807.0789}{{\tt arXiv:0807.0789}}].

\bibitem{Olv92}
{P. J. Olver}, {\it {Symmetry and explicit solutions of partial differential
  equations}},  {\em {Appl. Num. Math.}} {\bf {10}} ({1992}) {307--324}.

\bibitem{Olv86}
{P. J. Olver}, {\em {Applications of Lie Groups to Differential Equations}}.
\newblock {Springer-Verlag New York}, {New York}, {1986}.

\end{thebibliography}\endgroup
\bibliographystyle{myJHEP}  

\end{document}